\documentclass[aps,twocolumn,raggedbottom,prb,showpacs,nobalancelastpage,amssymb,groupedaddress]{revtex4}
\usepackage[latin1]{inputenc}
\usepackage[T1]{fontenc}
\usepackage[english]{babel}
\usepackage{graphicx}
\usepackage{amssymb}
\usepackage{amsmath}
\usepackage{amscd}
\usepackage{eucal}
\usepackage{color}
\usepackage{bm}
\usepackage{upgreek}

 \newcommand{\ignore}[1]{\relax}
 \def\det{\delta }

 \newcommand{\hf}{{1\over 2}}

 \DeclareMathAlphabet{\mathpzc}{OT1}{pzc}{m}{it}

\usepackage{epstopdf}
\def\le{\left}
\def\rg{\right}

\newcommand{\nc}{\newcommand}

\nc{\be}{\begin{equation}}
\nc{\ee}{\end{equation}}

\nc{\bea}{\begin{eqnarray}}
\nc{\eea}{\end{eqnarray}}
\nc{\bsub}{\begin{subequations}}
\nc{\esub}{\end{subequations}}
\nc{\nonu}{\nonumber}

 \DeclareMathOperator{\Tr}{Tr}

 \newcommand{\al}{\alpha}
 \newcommand{\bet}{\beta}

 \newcommand{\ga}{\gamma}
 \newcommand{\om}{\omega}
 
 \newcommand{\lam}{\lambda}

 \newcommand{\tEo}{\tau_{\rm E}^{\rm op}}
 \newcommand{\tEc}{\tau_{\rm E}^{\rm cl}}
  \newcommand{\tEe}{\tau_{\rm E}^{\rm e}}
 \newcommand{\tE}{\tau_{\rm E}}

 \newcommand{\tD}{\tau_{\rm D}}
 \newcommand{\tDI}{\tau_{\rm D1}}
 \newcommand{\tDII}{ \tau_{\rm D2}}
  \newcommand{\tH}{ \tau_{\rm H} }

  \newcommand{\Nn}{N_{\al}}
    \newcommand{\Nm}{N_{\bet}}

 \newcommand{\mS}{{\bf S}}

\nc{\hbom}{\hbar\omega}
\nc{\hfhbom}{{\hbar\omega\over 2}}

\DeclareMathOperator{\re}{\Re e}

\begin{document} 
\title{Semiclassical approach to the ac-conductance of  chaotic cavities}
 \author{Cyril Petitjean$^1$, Daniel Waltner$^1$, Jack Kuipers$^1$, \.{I}nan\c{c} Adagideli$^{1,2}$ and  Klaus Richter$^1$} 
 \address{$^1$Institut~f\"ur~Theoretische~Physik,~Universit\"at~Regensburg,~93040~Regensburg,~Germany.\\
$^2$Faculty~of~Engineering~and~Natural~Sciences,~Sabanci~University,~34956 Tuzla~Istanbul,~Turkey.} 
\date{\today}
 
 \begin{abstract}
We address frequency-dependent quantum transport through mesoscopic conductors in the semiclassical
limit. By generalizing the trajectory-based semiclassical theory of dc quantum transport to the ac case,
we derive the average screened conductance as well as ac weak-localization corrections for chaotic
conductors. Thereby we confirm respective random matrix results and generalize them by accounting
for Ehrenfest time effects. We consider the case of a cavity connected through many leads to a 
macroscopic circuit which contains ac-sources. In addition to the reservoir the cavity itself  is
capacitively coupled to a gate. By incorporating tunnel barriers between cavity and leads
we obtain results for arbitrary tunnel rates. Finally, based on our findings we investigate 
the effect of dephasing on the charge relaxation resistance 
of a mesoscopic capacitor in the linear low-frequency regime.
  \end{abstract}
 \maketitle

\section{Introduction} 

In contrast to  dc-transport experiments, the applied external frequency $\omega$  of an ac-driven
mesoscopic structure provides  a new energy scale $\hbar \omega$ that permits one to access further
properties of these systems, including their intrinsic charge distribution and dynamics.  
 
The interest in the ac-reponse of mesoscopic conductors goes back to the work of Pieper and
Price~\cite{Pie94a} on the dynamic conductance of a mesoscopic Aharonov-Bohm ring. This pioneering work
was followed by several experiments ranging from photon-assisted transport to quantum shot
noise~\cite{Che94,Kou94,Rez95,Ver95,Scho97,Rey03}. More recently, the ac-regime has been experimentally
reinvestigated achieving the measurement of the in and out of  phase parts of the
ac-conductance~\cite{Gab06a} and the realization of a high-frequency single electron
source~\cite{Fev07}.  Moreover, the recent rise of interest in the  full counting statistics of charge
transfer has led to a reexamination of the frequency noise spectra~\cite{Nag04,Hek06,Sal06}.  This
experimental progress has since triggered renewed  theoretical interest in time dependent mesoscopic
transport~\cite{ Bag07,Nig07,Mos08,Saf08,Par08}.

One way to tackle the ac-transport problem is to start from linear response theory for a given potential
distribution of the sample~\cite{Fis81,Bar89,She91}. This involves the difficulty that, in principle, the
potential distribution and more precisely its link to the screening is unknown. 
Another approach consists of deriving the ac-response to an external perturbation that only 
enters into quantities describing the reservoirs. Such approachs were  initiated  
by Pastawski~\cite{Pas91} within a non-equilibruium Green function based generalized Landauer-B\"uttiker formalism, and  then the scattering matrix formalism of a time-dependent system
was  developed by B\"uttiker et al.~\cite{But93a,But93c}. Since the energy is in general no longer conserved for an ac-bias, 
the formalism is based on the concept of a scattering matrix that depends on two energy 
arguments~\cite{Vav05} or equivalently on two times~\cite{Pol03}. Fortunately, when the inverse  frequency 
is small compared to the time to  escape the cavity, the ac-transport can be expressed in terms 
of the derivative of the  scattering matrix with respect to energy~\cite{But97}.  In this article we  
start from the time dependent  scattering matrix formalism and limit our investigations to 
open, classically chaotic ballistic conductors in the low-frequency regime~\cite{note1}.

For ac-transport we  calculate the average correlator of scattering matrices $\mS(E)$ at different energies $E$. For this we need to know the joint distribution of the matrix elements $S_{\al\bet;ij}$ at different values of the energy or other parameters. (We label the  reservoirs connected to the conductor by a greek  index  and the mode number by a latin index.)
  To our knowledge a general solution to this problem does not yet exist  for  chaotic systems.  However, in the limit of a large number of channels, the first moments of the distribution 
  $S_{\al\bet;ij}(E) S_{\al\bet;ij}^{\dagger}(E^{\prime})$ were derived using 
  both semiclassical methods~\cite{Blu88,Bar93a} and  
  various random matrix theory  (RMT) based methods~\cite{Ver85,Fra95, Brou97a,Pol03}.   
   Although the ac-transport properties of ballistic chaotic systems seem to be  well described by the RMT  of transport~\cite{Brou97a}, we develop a semiclassical approach for three  reasons:   First, this allows us to confirm the random matrix prediction by using a complementary trajectory-based  semiclassical method. Second,  the energy   dependence in the random matrix formalism was    introduced by resorting to artificial models such as the  "stub model"~\cite{Pol03}. 
   While being powerful, this treatment is far from microscopic or natural.  
   The  third and  strongest reason is to go beyond the  RMT treatment and investigate   the crossover
   to the classical limit.  Similarly as for the static case RMT is  not applicable in this regime.  As
   first noticed by Aleiner and Larkin~\cite{Ale96}, ballistic  transport is characterized by a new time   scale,  known as the  Ehrenfest time $\tE$~\cite{Lar69,Berm79},  that controls the appearance of interference effects.  The   Ehrenfest time  corresponds to the time during which a localized wavepacket spreads
   to a classical  length scale. Typically, in open chaotic systems two such lengths are relevant, the system size $L$ and the lead width $W$.  We can thus define an Ehrenfest time associated with each one~\cite{Vav03,Sch05}, the closed-cavity Ehrenfest time, 
 \be\label{tec}
 \tEc= \lambda^{-1} \ln[ L/\lambda_{\rm F}],
 \ee
and the open-cavity Ehrenfest time,
 \be\label{teo}
\tEo= \lambda^{-1} \ln[ W^2/\lambda_{\rm F}L],
\ee
where $\lambda$ is the classical Lyapunov exponent of the cavity.

Although the success of the semiclassical method (beyond the so-called diagonal approximation, see 
below) to describe quantitatively universal and non universal dc-transport properties is now clearly 
established~\cite{Walbook, Ric02, Ada03, Mul07, Jac06a, Brou06c, Brou07c, Whi07, Pet07c, Whi08, Kui08,Brou08}, 
the corresponding semiclassical understanding of frequency dependent transport is far less developed.
Based on an earlier semiclassical evaluation of matrix element sum rules by Wilkinson~\cite{Wil87}
and a semiclassical theory of linear response functions~\cite{Meh98}, a semiclassical approach to the 
frequency-dependent conductivity within the Kubo-formalism led to an expression of the ac-(magneto-) conductivity$\,$ $\sigma(\omega)$ in terms of a trace formula for classical periodic 
orbits \cite{Ric00}.  Closely related to this evaluation of $\sigma(\omega)$ is the problem of 
frequency-dependent (infrared-) absorption$\!$ in ballistic mesoscopic cavities which has been 
treated semiclassically in Ref.~[\onlinecite{Meh98}]. Peaks in the absorption could be assigned to 
resonance effects when the external frequency $\omega$ corresponds to the inverse periods of 
fundamental periodic orbits in the cavity. Ref.~[\onlinecite{Ale96}] contains a first, 
$\sigma$-model based approach to weak localization effects in the ac-Kubo conductivity, where the 
findings were interpreted in a quasiclassical trajectory picture (beyond the diagonal
approximation).  
 We note also  that the semiclassical treatment  of the product of scattering matrices $\mS(E)$ at different energies, 
 has been investigated in different context such as the    Ericson fluctuations~\cite{Mul07} and  the time delay~\cite{Kui08},  however without considering the Ehrenfest time dependence.

The outline of this article is as follows:  In Section~\ref{model}  we introduce our model to treat  the system of interest namely a quantum dot under ac bias,  and   recall some basic results about conservation laws in presence of a time dependent field.  In
Sect.~\ref{method}  we present the method used to treat screening, which  is based  on a self-consistent approach developed by B\"uttiker et al.~\cite{But93c}.  The  admittance, i.e.\  the ac-conductance, is  then  calculated semiclassically for the particular case of strong coupling to the leads (transparent contact) in Sect.~\ref{sec-notunel}, where we illustrate our result by treating the time dependence of a pulsed cavity.   We generalize the method to cope with  arbitrary  tunnel rates in Sect.~\ref{sec-tunel}, and finally we use our general results to investigate dephasing effects on the charge  relaxation resistance of a mesoscopic capacitor in Sect.~\ref{mesocapa}.

\section{The Model }\label{model}

\begin{figure}[h]
\begin{center}
\rotatebox{0}{\resizebox{8cm}{!}{\includegraphics{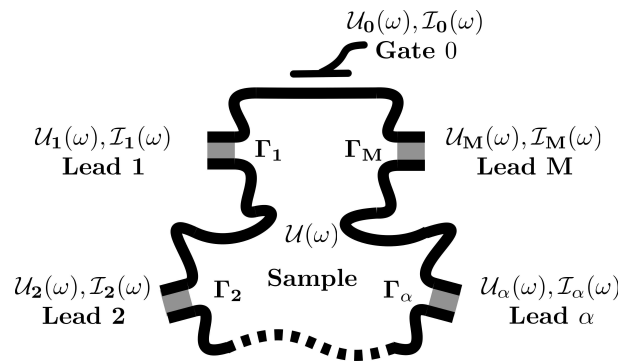}}}
\caption{Two dimensional chaotic cavity with $M$ leads and one gate $0$. Each   lead $\al$ has a width $W_{\al}$ and is coupled to a reservoir at potential ${\cal U}_{\al}(\om)$ and current ${\cal I}_{\al} (\om)$. Each  tunnel barrier is   characterized by the set of transmission probabilities ${\bf \Gamma}_{\al} =\{\Gamma_{\al,1}, \cdots,  \Gamma_{\al,N_\al} \}$. The  gate and the sample are capacitively coupled, which leads to a gate current ${\cal I}_{0} (\om) = - {\it i} \om C [{\cal U}_{0}(\om)-{\cal U}(\om)] $. }
\label{fig1}
\end{center}
\end{figure}

We consider a ballistic quantum dot, i.e.\ a two-dimensional chaotic cavity coupled to $M$ electron reservoirs via $M$ leads.  Each lead $\al$ has a width $W_\al$ and is coupled to the cavity through a tunnel  barrier (see Fig.~\ref{fig1}). In addition to the treatment of Ref.~[\onlinecite{Whi07}] we assign a particular tunnel probability to each lead mode. The  tunnel barrier is thus characterized by a set of  transmission probabilities, ${\bf \Gamma}_{\al} =\{\Gamma_{\al,1}, \cdots,  \Gamma_{\al,N_\al} \}$, with $N_\al$ the maximum mode number of lead $\al$. 
The chaotic dot is additionally  capacitively coupled to a gate connected  to a reservoir at voltage ${\cal U}_0(\omega)$, from which a current  ${\cal I}_0(\om)$ flows. This capacitive coupling with the gate is taken into account via a geometrical capacitance $C$~\cite{But93a,Gop96, Brou97a}.

We further require that the size of the contact is much smaller than the system size ${\rm L}$,
but still semiclassically large, $1 \ll N_{\al} \ll {\rm L}/\lambda_{\rm F}$. This requirement ensures that the particle spend enough time inside the cavity to experience the chaotic dynamics.

As  usual for such mesoscopic structures we need to distinguish between quantum and  classical time scales. 
On the quantum  side we have already introduced the Ehrenfest times ($\tEo$, $\tEc$) in Eqs.~(\ref{tec},\ref{teo}), while  another time scale is the Heisenberg time $\tH$, the time  to resolve  the mean level spacing of the system.
On the classical side the time of flight  $\tau_{\rm f}$  between two consecutive bounces at the system 
cavity wall is relevant. In  most ballistic systems or  billiards  we have $\tau_{\rm f} \simeq \lam^{-1}$.  
 Another relevant time scale is the  ballistic ergodic time  $\tau_{\rm erg}$  which determines  how long 
 it takes  for an electron to visit most of the  available phase space. However, as we deal with transport 
 properties, a further important time scale is the dwell time $\tD$, the average time spent in the cavity 
before reaching the contact, we have $\tD  /  \tau_{\rm erg}  \gg 1$.  The  related escape rate  therefore satisfies
\be
\label{tddef}
\tD^{-1} = \tH^{-1}   \sum_{\al=1}^{M} \sum_{i=1}^{N_{\al}} \Gamma_{\al,i}.
\ee
For small openings which  we consider here, we have  $\lam \,\tD \gg1 $. 

The ac-transport properties of such a mesoscopic system are characterized by the dimensionless  admittance 
\be
 g_{\al\bet}(\omega)  =
 G_{\al\bet}(\omega)/  G_{0}  = G^{-1}_{0} \partial I_{\al}(\omega) / \partial U_{\bet}(\omega),
 \ee
 with $G_0= d_s e^2/h$, where $d_s =1$ or $2$ in the absence or presence of spin degeneracy. 
 In this study we limit ourselves to the coefficients $g_{\al\bet}(\omega)$ with $\al,\bet =1,\cdots,M$ 
 where the coefficients denoting the gate are determined by current conservation and the freedom to 
 choose the zero point of energy~\cite{But93a}, 
 \be
 \label{current-cons}
 \sum_{\al=0}^M g_{\al\bet}(\omega) = \sum_{\bet=0}^M g_{\al\bet}(\omega) =0 \, .
\ee

We note that    Eq.~(\ref{current-cons}) is  a straightforward consequence of the underlying gauge invariance. 
Owing to the conservation of charge, the total electric current  fulfills the continuity equation
\be\label{continuity}
 {\pmb \nabla} \cdot {\bf j}_{p}   + {\partial  \rho \over \partial t}  = 0, 
 \ee
 where $\rho$ is the charge density and $ {\bf j}_{p}$ the particle current  density. For dc-transport, the charge density is time independent and so we have $ {\pmb \nabla} \cdot {\bf j}_{p} = 0 $.  Thus the sum of all currents that enter into the dot  is always zero. Moreover the  current properties must remain unchanged under a simultaneous global shift of the voltages of the reservoirs. These conditions  imply  the well know unitarity of the scattering matrix\cite{Blan00},
\be
\sum_{\al,i} S_{\al\bet;ij}^{\dagger}(E) S_{\al\ga;ik}(E)  = \delta_{\bet\ga;jk}.
\ee 
 For ac-transport,  the product of scattering matrices  at different energies no longer obey a similar property~\cite{Pre96,Wan97,Blan00,But00} i.e.\
\be
\sum_{\al,i} S_{\al\bet;ij}^{\dagger}(E) S_{\al\ga;ik}(E^{\prime}) \not = \delta_{\bet\ga;jk},
\ee
indeed this inequality expresses the fact that, due to the possible  temporary pile up of charge in the cavity,
 the particle current  density  no longer satisfies   $ {\pmb \nabla} \cdot {\bf j}_{p} = 0 $. However  one can instead use
 the Poisson equation 
  \be\label{poisson}
{\pmb \nabla} \cdot  {\bf D} =  \rho, 
\ee
where ${\bf D} = -\epsilon_0{\pmb \nabla} {\varphi} $ with $ {\varphi} $  the electric potential, 
to  define the total electric current density which satisfies  $ {\pmb \nabla} \cdot {\bf j} = 0 $, as a  sum of a particle and a displacement current:   
 \be\label{currentdens}
 {\bf j} = {\bf j}_{p} +  {\partial {\bf D} \over \partial t}.
  \ee
In order to find $ {\bf j} $ one needs to know the electrical field ${\bf D}$.  In general its calculation
is not a trivial task because  the intrinsic many-body aspect of the problem makes the treatment of the Poisson  equation (\ref{poisson}) tricky, especially if it is necessary to treat  the particle and displacement current on the  same footing.

In this work we shall adopt the approach of Ref.~[\onlinecite{But93c}] to simplify the problem. In this
approach the environment is reduced to a single gate, the Coulomb interaction is described by a
geometrical capacitance $C$, and the two currents are treated on different footing; the particle current is 
calculated quantum mechanically via the scattering approach, while the displacement current is treated
classically via the electrostatic law (Eqs.~(\ref{continuity},\ref{poisson})).  
This simplification will permit  us below to re-express the Poisson equation (\ref{poisson})  
to obtain the simplest gauge invariant
theory that takes care of the screening.  We  emphasize that even  though our model could be thought of
as oversimplified     it  has the advantage of being able  to probe the effects due to the long range
Coulomb interaction. Indeed, for non-interacting particles  it is possible  to treat the dot and  the
gate via two  sets  of uncorrelated continuity equations.  The  Coulomb interaction removes this
possibility, and  we need to consider the gate and dot  as a whole system. 

\section{Expression for the admittance}
\label{method}

 The method to compute the admittance  proceeds in two steps\cite{Pre96}:  First  the direct
 response (particle current) to the change of the external potential is calculated  under the
 assumption that the internal potential ${\cal U }(\omega) $ of the sample is fixed. This leads to the definition of the unscreened admittance $g_{\al\bet}^{u}(\omega)$.  Second,  a self-consistent procedure based on the gauge invariance  (current conservation and freedom to choose the zero of voltages)  is  used  to obtain the screened admittance $g_{\al\bet}(\omega)$.  
 
The unscreened admittance reads~\cite{But93a}
\bea\label{eq:unscreen}
&&g^{u}_{\al\bet}(\om)= \int\!{\rm d}E \frac{f(E-\hfhbom)- f(E+\hfhbom))}{\hbom}
\\ &&
\quad
\times
\Tr \le[ 
\delta_{\al\bet} {\bf 1}_{\al} -\mS_{\al\bet}\le(E+\hfhbom\rg)
\mS_{\al\bet}^{\dagger}\le(E-\hfhbom\rg)\rg],\qquad\nonu
\eea
where $f(E)$ stands for the Fermi distribution, $\mS_{\al\bet}$ is the $\Nn\times \Nm$ scattering 
matrix from lead $\bet$ to lead $\al$, and ${\bf 1}_{\al}$  is an $\Nn\times \Nn$ identity matrix. 
Under the assumption that ${\cal U }(\omega)$ is spatially uniform, the screened admittance $g_{\al\bet}(\omega)$ is  straightforward to obtain~\cite{But93a}. For sake of completeness  we present here only the outline of the method  and  refer  to  Ref.~[\onlinecite{But97}] for  more details.

On the one hand the current reponse at contact $\al$ is 
\be
{\cal I}_{\al}(\om) = G_{0}\le[ \sum_{\bet=1}^M g^{u}_{\al\bet}(\om) {\cal U}_{\bet}(\om) +  g^{i}_{\al0}(\om) {\cal U}(\om) \rg],
\ee 
\noindent where $g^{i}_{\al0}(\om)$ is the unknown internal reponse of the mesoscopic conductor generated by the  fluctuating potential ${\cal U}(\om) $.
 On the other hand the current induced at the gate is 
 \be
 {\cal I }_{0}(\om) = - {\it i} \om C [{\cal U}_0(\om) - {\cal U}(\om) ] .
 \ee
Gauge invariance permits a shift of $- {\cal U}(\om) $ and  provides an expression for the  unknown internal response,
\be 
g^{i}_{\al0}(\om) =  -  \sum_{\bet=1}^M g^{u}_{\al\bet}(\om).
\ee 
Then current conservation, $\sum_{\al=1}^M {\cal I}_{\al}(\om)  + {\cal I}_{0}(\om) =0$, yields the 
result of the screened admittance~\cite{But93a}, 
\be\label{eq:screen}
 g_{\al\bet} (\omega)= 
   g_{\al\bet}^{u} (\omega)+
   \frac{
   \sum_{\det=1}^M g_{\al\det}^{u} (\omega)
   \sum_{\det'=1}^M g_{\det'\bet}^{u} (\omega)
   }
   {
   {\it i}  \omega C/ G_0
   -
   \sum_{\det=1}^M\sum_{\det'=1}^M g_{\det\det'}^{u} (\omega)
   }.
\ee 
In the self-consistent approach used to obtain Eq.~(\ref{eq:screen}),  the only electron-electron interaction  
term that has been considered is the capacitive charging energy of the cavity. This implies  that we should  
consider a sufficiently large quantum dot~\cite{Ale02}.    We note  that, using a $1/N$-expansion,  the  self-consistent approach above was 
recently formally confirmed  in Ref.~[\onlinecite{Brou05b}].    Moreover,    Eq.~(\ref{eq:screen}) can be  generalized to non-equilibrium problems, using Keldysh non-equilibrium Green functions~\cite{Wan99}.  

In the next sections we  present the semiclassical evaluation of Eq.~(\ref{eq:unscreen}) in the zero
temperature limit (including finite temperature is straightforward). For reasons of presentation we
first give the semiclassical  derivation for the transparent case in Sect. \ref{sec-notunel}, and then we explore the general case in Sect.~\ref{sec-tunel}.   In Sect.~\ref{mesocapa}  we  present an application of the screened result for  tunnel coupling, when we compute   the relaxation resistance of a mesoscopic chaotic capacitor.

\section{Semiclassical theory for the admittance }\label{sec-notunel}
 
\subsection{Semiclassical approximation}
  
We first consider the multi-terminal case assuming  transparent barriers, i.e.\  
$\Gamma_{\al,i} = 1$,  $\forall (\al, i)$. In the limit  $k_{\rm B}T \to 0 $ 
the unscreened admittance, Eq.~(\ref{eq:unscreen}), reduces to
  \be\label{eq:unscreenT0}
g^{u}_{\al\bet}(\om)=\!
N_{\al}\delta_{\al\bet} \!- \!
\Tr \le[ 
\mS_{\al\bet}(E_{\rm F}\!+\!\hfhbom)\mS_{\al\bet}^{\dagger}(E_{\rm F}\!-\!\hfhbom)
\rg].\quad
\ee
 
Semiclassically,  the matrix elements for  scattering  processes from mode $i$ in lead $\bet$ to mode $j$ in lead $\al$ read~\cite{Mil75,Bar93a}
 \bea\label{sc-smatrix}
&&
\hspace{-0.5cm}
S_{\al\bet;ji} (E_{\rm F}\pm \hfhbom) = 
\\
&&-
\int_{\bet} \!\!\!{\rm d} {x}_0
\int_{\al} \!\!\!{\rm d} { x} 
{\langle  j \vert  { x} \rangle \langle  { x}_0 \vert  i \rangle \over 
(2 \pi {\it i} \hbar)^{1/ 2}
}
\sum_{\gamma} A_{\gamma}
e^{
 {{ \it i } \over \hbar} 
{ S}_{\gamma} ( { x}, { x}_0; E_{\rm F}\pm \hfhbom)
}
,
\nonumber 
\eea 
where $\vert i\rangle $ is  the transverse wave function of the $i$-th  mode. Here the $x_0$ (or $x$) integral is over the cross section of the $\bet$th (or $\al$th) lead. At this point $\mS_{\al\bet}$ is given by a sum over classical trajectories, labelled by $\gamma$. The classical paths $\gamma$ connect ${\bf X}_0 = (x_{0},p_{x_0})$ (on a cross section of lead $\bet$) to ${\bf X} = (x,p_{x})$ (on a cross section of lead $\al$).   
Each path gives a contribution oscillating with  action ${S}_{\gamma} $ (including  Maslov indices)
evaluated at the energy $E_{\rm F}\pm \hbar\om/2$ and weighted by the  the complex amplitude $A_\gamma
$.   This  reduces to the square root of an inverse element of the stability matrix~\cite{Gut90},
i.e.\ $A_\gamma  = \vert({\rm d}  p_{x_0} / {\rm d} x )_{\gamma} \vert^{\hf} $.

We insert Eq.~(\ref{sc-smatrix}) into Eq.~(\ref{eq:unscreenT0}) and obtain  double sums  over paths $\gamma$, $\gamma'$ and lead modes $\vert i\rangle$, $\vert j \rangle$.  The sum over the  channel indices  is  then performed with the semiclassical approximation~\cite{Whi07}, $\sum_{i=1}^{N_{\bet} } \langle x_{0} \vert i \rangle  \langle i \vert  x^{\prime}_{0}  \rangle\approx \delta(x^{\prime}_{0} -x_{0})$, and yields
 \be\label{eq:sc-unscreenT0}
g^{u}_{\al\bet}(\om) - N_{\al}\delta_{\al\bet} =
-\int_{\bet} \!\!\!{\rm d} {x}_0 \int_{\al} \!\!\!{\rm d} { x} 
\sum_{\gamma,\gamma'}  { A_{\gamma}A _{\gamma'}^{\ast} \over 2\pi\hbar} 
e^{{{ \it i } \over \hbar} 
\delta S(E_{\rm F} ,\om)
}.
\,
\ee
Here, 
\be\label{deltaS}
\delta S(E_{\rm F} ,\om)  = { S}_{\gamma} ( { x}_0, { x}; E_{\rm F}+ \hfhbom)
-{ S}_{\gamma'} ( { x}_0, { x}; E_{\rm F}-\hfhbom).
\ee
As we are interested in the limit $\hbar \om \ll E_{\rm F}$, we can  expand  $\delta S(E_{\rm F} ,\om)  $ around $E_{\rm F}$.  The dimensionless ac-conductance is then given by 
\bea\label{eq:sc-unscreenT0-ex}
g^{u}_{\al\bet}(\om) - N_{\al}\delta_{\al\bet} &=&
-\int_{\bet} \!\!\!{\rm d} {x}_0 \int_{\al} \!\!\!{\rm d} { x} 
\sum_{\gamma,\gamma'}
  { A_{\gamma}A _{\gamma'}^{\ast} \over 2\pi\hbar} 
 \\ &&\!\times
\exp\le[ 
{{ \it i } \over \hbar} 
\delta S(E_{\rm F})+ {{\it i} \om  \over 2}(t_{\gamma} +  t_{\gamma'})
\rg],
  \nonumber
\eea
\noindent where $\delta S(E_{\rm F}) = { S}_{\gamma} ( { x}_0, { x}; E_{\rm F}) - { S}_{\gamma'} ( { x}_0, { x}; E_{\rm F}) $  and 
$t_{\gamma}$ ($t_{\gamma'}$) is the total duration of the path $\gamma$ ($\gamma'$). Eq.~(\ref{eq:sc-unscreenT0-ex}) is the starting point of our further investigations.

\subsection{Drude Admittance}\label{sec-2Tnotunel}

  We are interested in quantities arising from averaging over variations in the energy or cavity shapes.
  For most sets of paths, the phase given by the  linearized action difference $\delta S(E_{\rm F})$
  will oscillate widely with these variations, so their  contributions will average out.  In the
  semiclassical limit, the dominant contribution  to Eq.~(\ref{eq:sc-unscreenT0-ex})  is the diagonal
  one, $\gamma =\gamma'$, which leads to $t_\gamma =t_{\gamma' }$, $\delta S(E_{\rm F}) =0$ and gives 
\bea\label{drude-cond2}
g_{\al\bet}^{u,{\rm D}} (\om)=
N_{\al}\delta_{\al\bet}
-\int_{\bet} \!\!\!{\rm d} {x}_0 \int_{\al} \!\!\!{\rm d} { x} 
\sum_{\gamma} \;
  { |A_{\gamma}|^2  \over 2\pi\hbar} 
 e^{{\it i} \om  t_{\gamma} }.
\eea
In the following we   proceed along the lines of Ref.~[\onlinecite{Jac06a}]. The key point is the
replacement of the semiclassical amplitudes by their corresponding classical probabilities. To
this end we use a  classical sum rule valid under  ergodic assumptions~\cite{Sie99b}, 
\bea\label{sumrule}
&&
\hspace{-0.5cm}
\sum_{\gamma} \;
   |A_{\gamma}|^2
 e^{{\it i} \om  t_{\gamma} }
 [\cdots]_\gamma
 =
\\
  &&\int_0^{\infty} \!{\rm d} t \int_{-\pi/2}^{\pi/2} \!\!\!{\rm d}\theta_0{\rm d}\theta\,
  e^{ {\it i} \om  t }
  p_{\rm F} \cos(\theta_0)
  {P}({\bf X}, {\bf X}_0;t)[\cdots]_{{\bf X}_0}.
\nonu
\eea 
In Eq.~(\ref{sumrule}),  $ p_{\rm F} \cos(\theta_0)$  is  the initial momentum along the injection lead and ${P}({\bf X}, {\bf X}_0;t)$ the classical probability density to go from an initial phase space point ${\bf X}_0 =(x_0,\theta_0)$ at the boundary between the system and the lead to the corresponding point  ${\bf X} =(x,\theta)$.  The average of ${P}$ over an ensemble or over energy gives a smooth function that reads
\be\label{classprob2}
\langle 
{P}({\bf X}, {\bf X}_0;t)
\rangle
=
{ \cos(\theta) \over 2 \tD
\sum_{\al=1}^{M}W_{\al}  } e^{-t/\tD} ,
\ee
with the escape rate $ \tD^{-1} $  given in Eq.~(\ref{tddef}).

Using Eqs.~(\ref{drude-cond2}), (\ref{sumrule}) and (\ref{classprob2}), we recover the Drude admittance
\be\label{drude-2}
g_{\al\bet}^{u,{\rm D} } (\om)=
N_{\al}\delta_{\al\bet}
-
{ N_{\al} N_{\bet} \over N} 
\le({1\over 1- {\it i} \omega \tD}\rg),
\ee
\noindent where  $N=\sum_{\al =1}^M N_{\al}$.

\subsection{Weak localization for transmission, reflection and coherent backscattering}

\subsubsection{Weak localization}

 \begin{figure}
 \includegraphics[width=8.5cm]{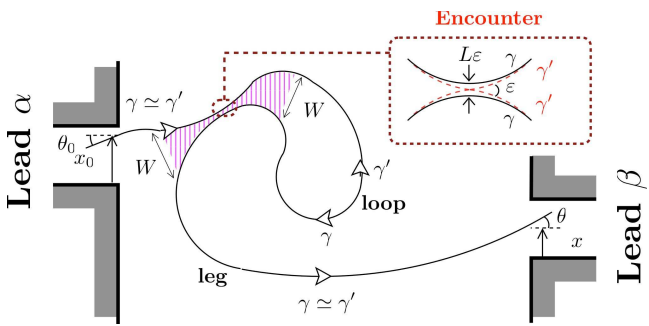} 
 \caption{\label{fig2} 
 A semiclassical contribution to weak localization for 
 a system with strong (transparent) coupling to the leads. The two paths follow each other closely
 everywhere except at the encounter, where one path (dashed line) crosses itself at an
 angle $\epsilon$, while the other one (full line) does not (going the opposite way around the loop).
 The cross-hatched area denotes the region where two segments of the solid paths are paired 
 (within $W_{\al} \simeq W_{\bet} \simeq W$ of each other) 
 }
 \end{figure}

The leading-order weak-localization correction to the conductance was identified in
Refs.~[\onlinecite{Ale96,Ric02}] as those arising from trajectories that are exponentially close almost
everywhere except in the vicinity of an encounter. An example of such a trajectory pair for chaotic
ballistic systems is shown in Fig.~\ref{fig2}.  At the encounter, separating the `loop' from the
`legs',
one of the trajectories ($\gamma'$) intersects itself, while the other one ($\gamma$) avoids the crossing. 
Thus, they travel along the loop they form in opposite directions. In the semiclassical limit, only pairs of trajectories with a small crossing angle $\epsilon$ contribute significantly to weak localization.
In this case, each trajectory remains correlated for some time on both sides of the encounter.
In other words, the smallness of $\epsilon$ requires two minimal times: $T_{\rm L}(\epsilon)$ to form a loop, and $T_{\rm W}(\epsilon)$ in order for the legs to separate
before escaping into different leads.  The encounter introduces a typical length scale $\delta r_{\perp}$ that corresponds to the perpendicular distance between the two paths in the vicinity of the encounter.  In the case of hyperbolic dynamics, we get $\delta r_{\perp} =v_{\rm F}\epsilon/(2\lambda)\sim L\epsilon $. Hence, 
the typical minimal time is given by $T_{\ell }(\epsilon) = \lambda^{-1} \ln [({\ell} /\delta r_{\perp})^2] $, with  ${\ell}  =\{{ \rm L, \, W } \}$ that we can approximate as
\begin{subequations}
  \begin{eqnarray}\label{TL}
T_{\rm L}(\epsilon) &  \simeq  & \lambda^{-1} \ln [\epsilon^{-2}], \\
\label{TW}
T_{\rm W}(\epsilon) & \simeq & \lambda^{-1} \ln [\epsilon^{-2} (W/L)^2].
\end{eqnarray}
\end{subequations}
The presence of the external driving does not change this picture. Each weak-localization contribution
accumulates a phase difference given  by the linearized action $\delta S(E_{\rm F}) \simeq \delta S_{\rm
RS} = E_{\rm F} \epsilon^2/\lambda$~\cite{Ric02}.  Following the same lines as for the derivation of the
Drude contribution, though the sum over  paths is now restricted to paths with an encounter, the sum rule (\ref{sumrule}) still applies, provided the probability ${P}({\bf X},{\bf X}_0;t)$ is restricted to paths which cross themselves.  To ensure this we write
\begin{eqnarray}
\!\!{P}({\bf X},{\bf X}_0;t)
\!&=& \!\! 
\int_{\cal C} {\rm d} {\bf R}_2 {\rm d} {\bf R}_1
{P}({\bf X},{\bf R}_2;t-t_2)
\nonumber \\
\!& \times & \!\!
{P}({\bf R}_2,{\bf R}_1;t_2-t_1)
{P}({\bf R}_1,{\bf X}_0;t_1) \,,\qquad
\end{eqnarray} 
where the integration is performed over the energy surface ${\cal C}$. Here, we use ${\bf R}_i=({\bf r}_i,\phi_i)$, $\phi_i \in [-\pi,\pi]$ for phase space points inside the cavity, while ${\bf X}$ lies on the lead surface as before.

 We then restrict the probabilities inside the integral to trajectories which cross themselves at phase space positions ${\bf R}_{1,2}$ with the first (or second) visit of the crossing occurring at time $t_1$ (or $t_2$).  We can write ${\rm d} {\bf R}_2 = v_{\rm F}^2 \sin \epsilon {\rm d}t_1 {\rm d}t_2{\rm d} \epsilon$ and set ${\bf R}_2 =({\bf r_1},\phi_1\pm \epsilon)$.  Then the weak-localization correction is given by 
\be \label{wl-2}
g_{\al\bet}^{u,{\rm wl}} (\om)= {1 \over \pi \hbar}\!\!
\int_{\bet} \! {\rm d} {\bf X}_0\! \int\! {\rm d} \epsilon
\re\le[ e^{{\it i} \delta S_{\rm RS} /\hbar} \rg]
\le\langle
F({\bf X}_0, \epsilon,\om)
\rg\rangle,
\ee
with,
\bea\label{F2}
&&\hspace{-0.5cm}
F({\bf X}_0, \epsilon,\om)
=
\\
&&
2v_{\rm F}^2 \sin\epsilon\int_{T_{\rm L}+ T_{\rm W}}^{\infty}
\!\!\! \!\!\!  {\rm d}t 
\int_{T_{\rm L}+ T_{\rm W}/2}^{t- T_{\rm W}/2}
\!\!\! \!\!\! {\rm d}t_2
 \int_{T_{\rm W}/2}^{t_2-T_{\rm L}}
 \!\!\! \!\!\!  {\rm d}t_1
\nonumber \\ 
&& \times
p_{\rm F} \cos \theta_0 
\int_{\rm R} {\rm d} {\bf Y} \int_{\cal C} \! {\rm d} {\bf R}_1 
{P}({\bf X},{\bf R}_2 ;t-t_2)  
\nonumber \\ 
&& \times
{P}({\bf R}_2,{\bf R}_1;t_2-t_1)  
{P}({\bf R}_1,{\bf X}_0;t_1)\,
e^{{\it i} \om t }.
\nonumber 
\eea

Under our approximation $t_{\gamma'} \simeq t_{\gamma } = t$,  the introduction of the driving frequency leads to performing a Fourier transform of the survival probability, and we obtain 
\bea\label{Fcal}
\le\langle
F({\bf X}_0, \epsilon,\om)
\rg\rangle
&=&
{ (v_{\rm F} \tD )^2 p_{\rm F}  \sin\epsilon \cos \theta_0 \over \pi \Omega}
{ N_{\al } \over N} 
\\&&\times
{ \exp\le[  - T_{\rm L} / \tD \rg]   \exp\le[ {\it i} \om (T_{\rm L} +T_{\rm W})\rg]
\over
(1-{\it i} \omega \tD)^3
},\nonumber
\eea
with $\Omega$ the cavity area. 
Inserting  Eq.~(\ref{Fcal}) into Eq.~(\ref{wl-2}), the $\epsilon$ integral is dominated by  small angle ($\epsilon \ll 1$)  contributions, allowing 
for  the approximation $\sin\epsilon \simeq \epsilon $ and pushing  the upper limit to infinity. This
yields  an  Euler Gamma function times an exponential term $e^{ -\tEc/\tD} e^{{\it i }\om (\tEc+\tEo)}$ (with $
\tEo $ and $\tEc$ given by Eqs.~(\ref{tec},\ref{teo})
that reads, to leading order in $(\lambda \,\tD)^{-1}$,
\bea\label{Gamma-wl-2}
&&
\hspace{-3mm}
\int_0^{\infty} 
{\rm d} \epsilon 
\,2
\re\le[
\exp\le[ {{\it i} E_{\rm F} \epsilon^2\over \lam \hbar}  \rg]
\rg]
\epsilon^{1+ {2\over \lam \tD}(1 -2{\it i} \om \tD)}
\le({{\rm W}\over {\rm L}}\rg)^{{ 2{\it i} \om\over\lam}}
\nonumber\\&&
\hspace{-3mm}
 \simeq
 - { \pi \hbar \over   m v_{\rm F}^2 \tD}  e^{ -{\tEc\over\tD}+{\it i }\om (\tEc+\tEo)}
  (1 -2{\it i} \om \tD) \!+
  \!{\cal O}\le[{1 \over \lambda \tD}\rg]
.\nonumber \\ &&\qquad
\eea
Performing the ${\bf X}_0$ integral and using $N_{\bet} = (\pi \hbar )^{-1} p_{\rm F} W_{\bet}$ and $N =(\hbar \tD)^{-1}m\Omega$, the weak-localization correction to the unscreened  admittance is 
\bea \label{uwl-2}
g_{\al\bet}^{u,{\rm wl}}(\om)
 =   \frac{N_{\al}  N_{\bet}}{N^2}  
e^{ -\tEc/\tD}
{(1-2{\it i}\om \tD) \, e^{{\it i }\om (\tEc+\tEo)}
\over
(1 - {\it i} \om \tD)^3
}.
\eea

We note that due to the absence of unitarity of the unscreened admittance we need to   explicitly
evaluate all the elements of $g_{\al\bet}^{u}(\om) $. The weak-localization contribution to reflection
$r_{\al\al}^{u,{\rm wl}} (\om)$ is derived in the same manner as $g_{\al\bet}^{u,{\rm wl}} (\om)$,
replacing however the factor $N_{\bet} /N$ by $N_{\al} /N$. We then obtain
\be \label{urwl-2}
r_{\al\al}^{u,{\rm wl}}(\om)
 =  \left( \frac{N_{\al}}{N} \right)^2
e^{ -\tEc/\tD}
{(1-2{\it i}\om \tD) \, e^{{\it i }\om (\tEc+\tEo)}
\over
(1 - {\it i} \om \tD)^3
}.
\ee

However  as in the dc-case  there is another leading-order  contribution to the reflection, the so-called coherent backscattering. This differs from weak localization as the path segments that hit the lead are correlated.  This mechanism should be  treated separately when computing the Ehrenfest time dependence, which is the object of the next paragraph.     
 
\subsubsection{Coherent backscattering}

Though the correlation between two paths does not influence the treatment of the external frequency, it
induces an action difference $\delta S(E_{\rm F}) = \delta S_{\rm cbs} =  -( p_{0\perp} + m \lambda
r_{0\perp}) r_{0\perp} $ where the perpendicular difference in position and momentum are $r_{0\perp} =
(x_{0} - x)\cos\theta_0$ and $p_{0\perp} = - p_{\rm F} (\theta -\theta_0)$.  As for weak localization,
we can identify two timescales, $\hf T'_{\rm L}, \hf T'_{\rm W}$, associated with the time for paths to
spread  to $L, W$, respectively.  However unlike for weak localization we define these timescales as
times measured from the lead rather than from the encounter.
Thus we have
\begin{eqnarray}
\label{eq:Tprimed}
T^{\prime}_{\ell }(r_{0\perp}, p_{0\perp}) \simeq  { 2 \over \lambda} 
\ln \le[ (m\lambda  {\ell} ) / \le\vert p_{0\perp} + m\lambda r_{0\perp}  \rg\vert\rg],
\end{eqnarray}
with ${\ell}  =\{ L,W \}$\cite{Whi08}.  Replacing the integral over ${\bf X}_0$ by an integral over $(r_{0\perp},p_{0\perp})$ and using $p_{\rm F}  \cos \theta_0 {\rm d} {\bf X}_{0} =  {\rm d}p_{0\perp}{\rm d}r_{0\perp}$, the coherent-backscattering contribution reads
\be \label{cbs-2}
r_{\al\al}^{u,{\rm cbs}} (\om)
= 
(\pi\hbar)^{-1}\!\!
\int_{\al} \!\! {\rm d}p_{0\perp}\!{\rm d}r_{0\perp}\!
\re\le[ e^{ { {\it i} \over \hbar }   \delta S_{\rm cbs} } \rg]\!\!
\le\langle
F^{\rm cbs}({\bf X}_0,\om)
\rg\rangle\!
,
\ee
with
\bea\label{Fcbs2}
\le\langle
F^{\rm cbs}({\bf X}_0,\om)
\rg\rangle
&=&
\int_{T'_{\rm L}}^{\infty} {\rm d}t 
\int_{\al } {\rm d} {\bf X} 
\;
{P}({\bf X},{\bf X}_0 ;t)  
e^{{\it i} \om t }
\nonumber \\
&=&
{N_{\al} \over  N}
{e^{- (T'_{\rm L}  - \hf T'_{\rm W}) / \tD}
e^{{\it i } \om T'_{\rm L} }
\over
1- {\it i}\om \tD
}.\qquad
\eea
As in the dc-case we perform a change of variables $\tilde{p} _{0\perp} =  p_{0\perp} +m\lam
r_{0\perp}$. Then we push the  $\tilde{p}_{0\perp}$ integral limit to infinity and evaluate the
$r_{0\perp}$ integral over $W_{\al}$. This result,
\bea
&&\int_{-\infty}^{\infty} {\rm d} \tilde{p}_{0\perp}
{\hbar \sin(\tilde{p}_{0\perp} W_{\al} /\hbar) \over \tilde{p}_{0\perp}} 
\le\vert { \tilde{p}_{0\perp} \over m\lam L}\rg\vert^{{ ( 1 - 2{\it i} \om\tD)\over \lam\tD}}
\le( {W\over L} \rg)^{{1\over \lam\tD}}
\nonumber \\
&&=
\pi\hbar \; e^{ -{\tEc\over \tD}}e^{{\it i }\om (\tEc+\tEo)}
+  {\cal O} \left[( \lambda \tD )^{-1} \right],
\eea
together with Eq.~(\ref{Fcbs2}) and Eq.~(\ref{cbs-2}) yields
 \bea \label{ucbs-2}
r_{\al\al}^{u,{\rm cbs}} (\om)
= -  \frac{N_{\al} }{N}  
e^{ -\tEc/\tD}
{ e^{{\it i }\om (\tEc+\tEo)}
\over
(1 - {\it i} \om \tD)
}.
\eea
 
Surprisingly the coherent-backscattering contribution   thus has exactly the same exponential dependence on $\tEo$ and $\tEc$ as the other weak-localization contributions. While in the dc-case this property is a consequence of current conservation, this fact is not obvious in the ac-case. 

At  this point we can  summarize our  results for the unscreened admittance. From Eqs.~(\ref{drude-2}, \ref{uwl-2}, \ref{urwl-2}, \ref{ucbs-2}), $\left\langle g^{u}_{\al\bet}(\om) \right\rangle$  can be written as 
   \begin{widetext}
\begin{equation}\label{unscreen-sc}
\left\langle  g_{\al\bet}^{u}(\omega)\right\rangle 
=
\delta_{\al\bet} N_{\al} 
-
\frac{N_{\al}N_{\bet}}{ N (1 - {\it i} \omega \tD)}
+
\frac{N_{\al}\, \exp\left[ -\frac{\tEc}{\tD}\right]
\,\exp\left[ {\it i } \omega(\tEc+\tEo)\right]}{N (1 - {\it i} \omega \tD)}
\left(
\frac{ N_{\bet} (1- 2 {\it i} \omega \tD )}{N (1 - {\it i} \omega \tD)^2}
- \delta_{\al\bet} 
\right)
+
{\cal O} (N^{-1}).
\end{equation}
  \end{widetext}

First we note that in the limit of zero Ehrenfest time we recover the RMT result  for the  unscreened admittance of Brouwer and B\"uttiker~\cite{Brou97a}.  Concerning the Ehrenfest time dependence of the admittance, we note that the result is consistent with the absorption study performed in Ref.~[\onlinecite{caveat2}].  As for the dc-case we find  the absence of the  Ehrenfest time $\tEo$ in the term $\exp[-\tEc/\tD] $
which derives from the classical correlation between the paths that constitute the encounter.  The  physical origin  of the term  $\exp\left[ {\it i } \omega(\tEc+\tEo)\right]$ comes  from the fact that both trajectories that contribute to weak localization and coherent backscattering involve an encounter that has a minimal duration of $(\tEc+\tEo)$  (Leg part and loop part of the encounter, see Fig.~\ref{fig2}).  The  presence of this minimal duration, $2\tEe = \tEc+\tEo$,  is in accordance with the Ehrenfest time shift prediction of the quantum correction to the survival probability~\cite{Wal08} and  the photofragmentation statistics~\cite{Gut08}.   We return to the Ehrenfest time dependence in Sect.~\ref{cavpulsed}.

We  can also consider the effect of a magnetic flux on the mesoscopic admittance. 
A weak magnetic field has little effect on the classical dynamics but  generates 
a phase difference between two trajectories that travel in opposite directions around 
a weak-localization generating  closed loop. This phase difference is $ \Phi / \Phi_0 $,
where $\Phi_0$ is the flux quantum, and $\Phi$ is proportional to  the flux through  
the directed area enclosed by the loop.  To incorporate this in the previous semiclassical 
treatment we must introduce a factor $\exp [{\it i}  \Phi / \Phi_0 ]$ into 
$F$ in Eq.~(\ref{Fcal}) and $F^{\rm cbs}$ in  Eq.~(\ref{Fcbs2}).  The calculation
gives a Lorentzian shape~\cite{Bar93a,Ric02,Jac06a} for the $ \Phi$-dependence of the 
quantum correction to the average admittance,
\be\label{gwlmagneto}
g^{u,{\rm wl}/{\rm cbs}}_{\al\bet}(\omega , { \Phi} )  = 
\frac{
g^{u,{\rm wl}/{\rm cbs}}_{\al\bet}(\omega ,0 )
}{
1 +  {\cal A}^2 { \Phi}^2   \le( \tau_{\rm f} / \tD - {\it i} \omega \tau_{\rm f} \rg)^{-1}
}.
\ee
Here ${\cal A}^2 =  \alpha  \Omega^2$, with $\alpha$ a system dependent parameter of order 
unity, $\Omega$ the cavity area and  $\tau_{\rm f}$ is the time of flight between two 
consecutive bounces at the cavity wall. 
 
 \subsection{The  screened  admittance}\label{wlocdep}

Following the  self-consistent approach, the screened admittance is straightforwardly obtained  when we  substitute Eq.~(\ref{unscreen-sc})  into Eq.~(\ref{eq:screen}) and expand the result to leading order in $N^{-1}$.  This simple substitution is justified, because the typical fluctuations of the unscreened admittance are of order $N^{-2}$.  The screened admittance then reads 
\begin{widetext}
 \bea\label{screen-sc}
\left\langle  g_{\al\bet}(\omega)\right\rangle 
=
\delta_{\al\bet} N_{\al} 
-
\frac{N_{\al}N_{\bet}}{ N (1 - {\it i} \omega \tau)}
+
\frac{N_{\al}\, \exp\left[ -\frac{\tEc}{\tD}\right]
\,\exp\left[ {\it i } \omega(\tEc+\tEo)\right]}{N (1 - {\it i} \omega \tD)}
\left(
\frac{ N_{\bet} (1- 2 {\it i} \omega \tau )}{N (1 - {\it i} \omega \tau)^2}
- \delta_{\al\bet} 
\right)
+
{\cal O} (N^{-1}),
\eea
\end{widetext}
where $\tau^{-1} = \tD^{-1} + NG_0/C$ is the charge relaxation time or quantum RC time.  
Eq.~(\ref{screen-sc}) is  the first intermediate 
result from which we can draw some general conclusions.
At zero Ehrenfest time we recover the two-terminal result of Brouwer and B\"uttiker in
Ref.~[\onlinecite{Brou97a}]. The comparison between the screened (Eq.~(\ref{screen-sc})) and  unscreened
(Eq.~(\ref{unscreen-sc})) admittance shows that the screening  amounts to the replacement  of the
dwell time $\tD$ by the RC time $\tau$ everywhere up to the prefactor of the third term.   Only
for the weak localization and the coherent-backscattering contributions does the dwell time
dependence survive. Though the relevant time scale for the classical admittance is the charge
relaxation time $\tau$, the quantum corrections are characterized by the dwell time $\tD$.  It is
important to remember that $\tD$ is a characteristic time scale of the non-interacting system.
Its relevance here has its origin in the fact that weak localization  is due to the interference
of electronic waves, which is unimportant for charge accumulation in the system. The absence of the RC time $\tau$ at  leading order in $\om$ is thus quite natural.   
We recall  that, as  constructed in the framework of the  model, the admittance matrix
Eq.~(\ref{screen-sc}) is current conserving if the gate is included.  The elements of the admittance
related to the gate are obtained via the sum rule (\ref{current-cons}).
Nevertheless, if we impose  this above sum  rule to the unscreened result  we   also  obtain a  conserved  current,  and this situation corresponds to a cavity which has  infinite capacitance to the gate.   In the reverse limit of zero capacitance we reach the charge neutral regime that corresponds to putting  $\tau=0$ in Eq.~(\ref{screen-sc}). Upon performing that, we recover the charge-neutral limit obtained by Aleiner and Larkin in Refs.~[\onlinecite{Ale96},\onlinecite{caveat3}] which for the conventional weak-localization contribution reads 
\be\label{screen-sc-CN}
 g^{{\rm wl}, \tau= 0}_{\al\bet}(\omega)
=
\frac{N_{\al} N_{\bet} } {N^2}
\,
\frac{
 \exp\le[-\frac{\tEc}{\tD} + {\it i } \omega(\tEc+\tEo)\rg] 
 }{ (1 - {\it i} \omega \tD)}.
\ee

We note that for the particular geometry  of a capacitor  (only one  lead and one  gate),  since  Eq.~(\ref{screen-sc})  is  valid for any capacitive coupling, we can obtain  the effect of the Ehrenfest  time scale  on  the interference correction to  the admittance of a mesoscopic capacitor.  This was not possible within the charge-neutral limit   approach of  Aleiner and Larkin, since the interference  corrections considered here are absent  in that case.

Here one important remark is due. In both, Eq.~(\ref{screen-sc}) and Eq.~(\ref{screen-sc-CN})  the admittance involves an oscillatory behavior as a function of the Ehrenfest time,   which should  in  principle be  more  easily accessible experimentally. Indeed, we see here in our quest for  the Ehrenfest time physics a clear advantage in investigating weak localization in the ac-regime. In the static case,  the ratio $\tE/\tD$ is  the only relevant and tunable parameter for the dc weak-localization correction. Consequently, the range of experimental investigation is   considerably reduced by the logarithmic dependence of  $\tE$ on  the system size.  For  the dynamical weak localization the frequency dependence $\om$ combined with the capacitive coupling $C$ provides more freedom in probing $\tE$-behavior.  
   However, although the $\om\tE$ Ehrenfest time dependence was predicted  in Ref.~[\onlinecite{Ale96}]
   (in which some possible experimental verification  was forecasted in a magnetocondutance experiment or
   in an optical backscattering experiment), we are not  aware of any experimental verification of the
   existence of such an oscillation. To date there exist only two experiments devoted to exploring the
   $\tE$ signature: The shot noise experiment by Oberholzer et al.\cite{Obe02} and the weak localization
   experiment in an antidot lattice by Yevtushenko et al.~\cite{Yev00}.  Both experiments were performed
 in the static case.

\subsection{Pulsed cavities}\label{cavpulsed}

In this section we comment on  the Ehrenfest time  dependence of  the admittance and  its link to that of  the survival probability \cite{Wal08,Gut08}. 
 To this end   we consider the particular case of a pulsed cavity~\cite{But00}, i.e.\ the application of a pulse ${\cal U}_{\al}(t) = a_{\al} \delta(t)$ to one of the contacts $\al$.  The response current at contact $\bet$ to  such a pulse will be  proportional to the frequency integral   over the ac-conductance,  
 \be
 g_{\al\bet}^{u}(t) ={1 \over 2\pi} \! \int \!{\rm d}\om \,g_{\al\bet}^{u}(\omega) \exp\le(-{\it i} \om t\rg).
 \ee  
  This problem was previously addressed in Ref.~[\onlinecite{But00}]  where the connection  between
  the RMT calculation of the admittance and   RMT results for the   quantum  and  the
  classical survival probability~\cite{Cas97, Fra97} were discussed. More precisely, in
  Refs.~[\onlinecite{Cas97, Fra97}]  a difference between the quantum  and  the classical survival
  probability was predicted for times of order  $t^{\ast}  = \sqrt{\tD \tH} $. The  conclusion of
  Ref.~[\onlinecite{But00}]  was two-fold:
 first, based on the weak-localization correction, a deviation of the  unscreened admittance
 at $t^{\ast}$ was confirmed,  while secondly the  screened system was shown not to exhibit
 such a $t^{\ast}$-dependence. 
  
Based on our semiclassical results~(\ref{unscreen-sc},\ref{screen-sc}) we are able to confirm 
this dependence. For the  unscreened  admittance, the weak-localization  and coherent-backscattering contribution, $\delta g_{\al\bet}^{u}(t)=  g_{\al\bet}^{u,{\rm wl}}(t) + g_{\al\bet}^{u,{\rm cbs}}(t)$, yields   a complicated time-dependence and reads on a log  scale  
  \bea\label{timecorrect}
  && 
  \hspace{-0.5cm}
  \ln\le[ { N\tD \over N_{\al}N_{\bet}}\,\delta g_{\al\bet}^{u}(t) \rg] 
=  - {t  - \tEo \over \tD} 
\\&&
  \hspace{1cm}
  +
\ln\le[
-{\delta_{\al\bet}\over N_{\al} }
+
{1\over N }
\le(
 {t  -2\tEe \over \tD} 
 \rg)
 \le(
 2 -
  {t  -2\tEe \over 2\tD} 
 \rg)
\rg]  . \nonumber
  \eea
Here we recall that  $2\tEe = \tEc+\tEo$. 
 At zero Ehrenfest time, $\tEe =0$, we see as in Ref.~[\onlinecite{But00}] that 
while the initial time dependence is  determined by $\tD$ (first term of rhs of Eq.(\ref{timecorrect})),
for times larger than $t^{\ast} $ the $t^2$-term in the log will be important. We therefore find 
a deviation from the classical  exponential behavior.

This conclusion  still holds at finite Ehrenfest time, up to the inclusion of  a time shift $2\tEe $ as
predicted in the recent semiclassical derivation\cite{Wal08} of the  survival probability.

The  treatment  of  the screened  case is more demanding  due to the presence of the RC time $\tau$. However since the pole linked  to the  dwell time $\tD$ is  only simple, it is clear that even at incomplete  screening, there is no term proportional to $t^2$. This is in accordance with the absence of deviations for the interacting  admittance. However, the  Ehrenfest time dependence  will be equivalent 
to the unscreened one, leading to a time shift .  Only for complete screening ($\tau = 0$) it is 
possible to obtain a simple result,  which reads on a log  scale 
 \bea
  && 
\hspace{-0.5cm}
  \ln\le[  {N\tD\over N_{\al}N_{\bet} } \,\delta g_{\al\bet}^{ \tau= 0}(t) \rg] 
=  - {t  - \tEo \over \tD} 
  +
\ln\le[
\frac{1}{N}
-{\delta_{\al\bet}\over N_{\al} }
\rg]\!.
\qquad
\eea

\section{Multi-terminal system with tunnel barrier}\label{sec-tunel}

The calculation of the admittance with tunnel barriers follows the trajectory-based method recently
developed by Whitney \cite{Whi07} for the dc-case. We recall here the  three main  changes in the theory with respect to the transparent case. For more details on the inclusion of tunnel barriers we refer to Ref.~[\onlinecite{Whi07}].

At first, in the presence of tunnel barriers  the complex amplitude $A_\gamma $ in Eq.~(\ref{sc-smatrix}) is extended to include the tunneling probabilities reading\cite{Whi07},
\bea\label{sc-amp}
A_\gamma &=& C^{\hf}_\gamma \,  t_{\bet,i}t_{\al,j}\prod_{\bet',j'} \le[ r_{\bet',j'}\rg]^{{\cal N}_{\gamma}(\bet',j')}
\eea
where $C_\gamma  = \vert({\rm d}  p_{x_0} / {\rm d} x )_{\gamma} \vert $  is the rate of change of the
initial momentum $p_{x_0}$ for the exit position $x$ of $\gamma$, ${\cal N}_{\gamma}(\bet',j')$ is the number of times that $\gamma$ is reflected back into the system from the tunnel barrier on lead $\bet'$ and the transmission and refection amplitudes at the lead $\bet$  satisfy $|t_{\bet,i}|^2 =(1 -|r_{\bet,i}|^2) =\Gamma_{\bet,i} $.  We note that without any loss of generality, we associated in Eq.~(\ref{sc-amp}) the momentum $ p_{x_0}$ (or $ p_{x}$) with  the channel $i$ (or $j$). 

At this point  the replacement of the semiclassical amplitudes by their corresponding classical
probabilities still holds, though the tunneling probabilities are included.  As an 
example the probability to go from a phase point ${\bf X}_0$ (here we associate the channel $i$ to the momentum $p_{\rm F} \cos\theta_0 $) on lead $\bet$ to an arbitrary point on lead $\al$ simply satisfies (for $\al\not=\bet$), 
 \bea
 \int_0^{\infty} \!
 {\rm d} t \int_{\al} {\rm d} {\bf X}  \le\langle {P}({\bf X},{\bf X}_0;t)\rg \rangle =
 {
 \Gamma_{\bet,i} \Gamma_{\al}^{(1)}
 \over 
 {\mathcal N}
 },
 \eea
\noindent where we let $\Gamma_{\bet}^{(1)}  =  \sum_{j=1}^{N_{\bet}}  \Gamma_{\bet,j}$ and define ${\mathcal N}= \sum_{\al}\Gamma_{\al}^{(1)}$.

More importantly, the introduction of a tunnel barrier induces three changes:
(i)
The dwell time (single path survival time) becomes
\begin{equation}
\tDI^{-1} = \tH^{-1} \sum_{\al} \Gamma_{\al}^{(1)} = \tH^{-1} {\cal N},
\end{equation}
because a typical path may hit a lead but be reflected off the tunnel barrier
(remaining in the cavity) numerous times before tunneling and  escaping.

(ii)
The paired-paths survival time for paths closer than the lead width
is no longer equal to the dwell time instead it is given by
\bea
\tDII^{-1} & = &  \tH^{-1} \sum_{\al}  \le ( 2 \Gamma_{\al}^{(1)} -\Gamma_{\al}^{(2)} \rg) \nonumber  \\
      & = &  \tH^{-1} \le ( 2 {\cal N} -\tilde{\cal N} \rg) ,
\eea
\noindent where $\Gamma_{\al}^{(2)}  =  \sum_{i=1}^{N_{\al}}  \Gamma_{\al,i}^2$ and we define
 $\tilde{\cal N}= \sum_{\al}\Gamma_{\al}^{(2)}$. This is because a second path following a path which has not escaped will hit the same tunnel barrier, and thus may escape even though the first path did not. Compare this with a system without 
 tunnel barriers: there a path has not escaped because it has not touched the leads; 
 thus a second path following the first one has no possibility to escape.

\begin{figure}
\begin{center}
\includegraphics[width=8.5cm]{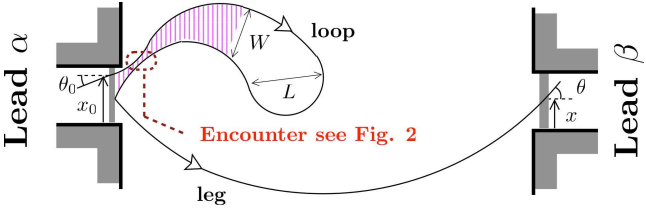} 
\caption{\label{fig3} 
A failed coherent-backscattering contribution to
ac-conductance, $g^{u,{\rm cbs}}_{\al\bet} (\om)$.
It involves paths which return close but 
anti-parallel to themselves at lead $\al$,
but  are reflected off the tunnel-barrier, remaining in the cavity 
to finally escape via lead $\bet$. The cross-hatched area denotes the region where 
the two solid paths are paired (within $W_{\al} \simeq W$ of each other).}
\end{center}
\end{figure}

(iii)
The coherent backscattering peak contributes to transmission as well as reflection.  The positive contribution to the transmission  competes  with  the usual negative weak-localization contribution to transmission, see also Fig~\ref{fig3}.
 
For the calculation of  the Drude conductance, only change (i) above is required, yielding 
\bea\label{ac-ud}
 g^{u,{\rm D}}_{\al\bet}  (\om)
 &=& \Gamma_{\bet}^{(1)}  \delta_{\al\bet}
 -
  {
 \Gamma_{\al}^{(1)} \Gamma_{\bet}^{(1)}
   \over
    {\mathcal N} 
  }  \frac{1}{1- {\it i} \omega \tDI } 
\eea
 
When calculating the conventional weak-localization contribution we need changes (i) and (ii) above.  Since the classical paths considered stay close to itself for a time $T_W(\epsilon)/2$ on either side of the encounter  we must use the paired-paths survival time, $\tDII$, for these parts of the path.  Elsewhere the escape time is given by the single path survival time, $\tDI$. With these new ingredients
we find that the conventional weak-localization contribution becomes
\be\label{ac-uwl}
 g^{u,{\rm wl}}_{\al\bet}  (\om)
 =
{\Gamma_{\al}^{(1)}  \Gamma_{\bet}^{(1)}   \over {\cal N}^2} 
    \frac{
    \le( 2 - {\tilde{\cal N} \over {\cal N} } \rg) 
    - 2{\it i}  \omega \tDI
    }{
    (1- {\it i} \omega \tDI)^3 
    } 
  e^{-\Theta_{\tE}}
    e^{{\it i} \omega (\tEc+\tEo) }, 
\ee
with $\Theta_{\tE} =  \tEo/\tDII+ (\tEc -\tEo)/\tDI $.  The exponential suppression $\exp(-\Theta_{\tE} )$ related to the classical correlation is simply the probability that the path segments survive a time $\tEo$ as a pair ($\tEo/2$ on either side of the crossing) and survive an additional time  $(\tEc-\tEo)$ unpaired (to complete a loop of length $\tEc$).  Similarly as for the transparent case, the exponential dependence $\exp[{\it i }\om(\tEc+\tEo)]$ indicates that the minimal duration of a weak localization trajectory is $ \tEc+\tEo$.

However as realized by Whitney~\cite{Whi07},  this is not the total weak-localization contribution to conductance, because of failed coherent-backscattering $ g^{u,{\rm cbs}}(\om)$ that contributes to conductance (change (iii) above). We recall that this involves a path which returns close but anti-parallel to itself at  lead $\al$, but is then reflected off the tunnel-barrier on lead $\al$, remaining in the cavity until it eventually escapes through lead $\bet$.  An example of such a trajectory  is shown in Fig.~\ref{fig3}.
We can calculate the backscattering contribution as before but using $\tDII$, when the paths are within $W_{\al}$ of each other, and  $\tDI$ elsewhere.  This result is then multiplied by the probability that the path reflects off lead $\al$ and then escapes through lead $
\bet $ and weighted by the dynamical factor $(1- {\it i} \omega \tDI )^{-1}$  due to the diagonal transmission from $\al$ to $\bet$ i.e.\ the leg part of Fig~\ref{fig3}.  In addition to the coherent backscattering expression for $r^{u,{\rm cbs}}(\om)$  this gives a contribution to the admittance of the form
 \begin{subequations}\label{ucbsf}
\begin{eqnarray}
 g^{u,{\rm cbs1}}_{\al\bet} (\om)
       &=&  
        \frac{\Gamma_{\al}^{(2)} - \Gamma_{\al}^{(1)}}{(1- {\it i} \omega \tDI)^2} \frac{\Gamma_{\bet}^{(1)}}{{\cal N}^2}
        e^{-\Theta_{\tE}}    e^{{\it i} \omega (\tEc+\tEo) }, \quad
         \\
g^{u,{\rm cbs2}}_{\al\bet}  (\om)
       &=&   
       \frac{ \Gamma_{\bet}^{(2)} - \Gamma_{\bet}^{(1)}}{(1- {\it i} \omega \tDI)^2}  \frac{\Gamma_{\al}^{(1)}}{{\cal N}^2}
       e^{-\Theta_{\tE}}    e^{{\it i} \omega (\tEc+\tEo) }, \quad \\ 
       r^{u,{\rm cbs}}_{\al\bet}  (\om)
  &=&  - 
    \frac{\delta _{\al\bet}   }{1- {\it i} \omega \tDI } 
    {\Gamma_{\al}^{(2)}  \over {\cal N}}
 e^{-\Theta_{\tE}}
    e^{{\it i} \omega (\tEc+\tEo) }, \qquad
\end{eqnarray}
\end{subequations}
\noindent where we recall that  $\Gamma_{\al}^{(2)}  =  \sum_{i=1}^{N_{\al}}  \Gamma_{\al,i}^2$. 

Using  Eqs.~(\ref{ac-ud}, \ref{ac-uwl}, \ref{ucbsf}), the unscreened admittance in the presence of tunnel barriers reads
   \begin{widetext}
\bea\label{unscreen-scM}
\left\langle  g_{\al\bet}^{u}(\omega)\right\rangle 
&=&
\Gamma_{\al}^{(1)}
\delta_{\al\bet} 
-
\frac{ \Gamma_{\al}^{(1)}  \Gamma_{\bet }^{(1)} }{ {\mathcal N} (1 - {\it i} \omega\tDI)}
 \\
 &&\!\!\!\!\!\!\!+
\frac{
\Gamma_{\al}^{(1)}\Gamma_{\bet}^{(1)} 
}{
{\mathcal N} ^2
}
\frac{
e^{-\Theta_{\tE}} e^{  {\it i} \omega (\tEc+\tEo) }
  }{
 (1 - {\it i} \omega\tDI)  
 }
\left(
\frac{
 2 - {\tilde{\cal N} /{\cal N} } 
 - 2 {\it i} \omega\tDI 
 }{
   (1 - {\it i} \omega\tDI)^2   
   }
 + 
 {
{\Gamma_{\al}^{(2)} / \Gamma_{\al}^{(1)}  }   +
  {  \Gamma_{\bet}^{(2)} /  \Gamma_{\bet}^{(1)}  }  -2
  \over   
  (1 - {\it i} \omega\tDI) }
-  
{ \Gamma_{\al}^{(2)} \over  \Gamma_{\al}^{(1)}}
{ {\cal N}  \over \Gamma_{\bet}^{(1)}  }\delta_{\al\bet} 
\right)
+
{\cal O} \le( N^{-1}\rg).\nonu
\eea
\end{widetext}

As a check of the formula~(\ref{unscreen-scM}), we can easily recover the previous 
 Eq.~(\ref{unscreen-sc}) for the unscreened admittance obtained for transparent barriers and also  the tunnel dc-conductance~\cite{Whi07}.
 
After the substitution of   Eq.~(\ref{unscreen-scM})  into Eq.~(\ref{eq:screen})  the screened admittance in presence of tunnel barriers reads
  \begin{widetext}
\bea\label{screen-scM}
\left\langle  g_{\al\bet}(\omega)\right\rangle 
&=&
\Gamma_{\al}^{(1)}
\delta_{\al\bet} 
-
\frac{ \Gamma_{\al}^{(1)}  \Gamma_{\bet }^{(1)} }{ {\mathcal N} (1 - {\it i} \omega\tau)}
\\
 &&\!\!\!\!\!\!\!+
\frac{
\Gamma_{\al}^{(1)}\Gamma_{\bet}^{(1)} 
}{
{\mathcal N} ^2
}
\frac{
e^{-\Theta_{\tE}} e^{  {\it i} \omega (\tEc+\tEo) }
  }{
 (1 - {\it i} \omega\tDI)  
 }
\left(
\frac{
  2 - {\tilde{\cal N} / {\cal N} } 
 - 2 {\it i} \omega \tau 
 }{
   (1 - {\it i} \omega \tau)^2   
   }
 + 
 {
{\Gamma_{\al}^{(2)}  / \Gamma_{\al}^{(1)}  }   +
  {  \Gamma_{\bet}^{(2)} /  \Gamma_{\bet}^{(1)}  }  -2
     \over
        (1 - {\it i} \omega \tau) }
-  
{ \Gamma_{\al}^{(2)} \over  \Gamma_{\al}^{(1)}}
{ {\cal N}  \over \Gamma_{\bet}^{(1)}  }\delta_{\al\bet} 
\right)
+
{\cal O} \le( N^{-1}\rg), \nonu
\eea
\end{widetext}
where the quantum RC time reads now  $\tau^{-1} = \tDI^{-1} + {\mathcal N} G_0/C$.  We emphasize that from  Eq.~(\ref{screen-scM}) it is possible  to derive all the results presented in this paper and therefore this equation is  the central result of this paper. 

In the second line of  Eq.~(\ref{screen-scM}),  the  second contribution  in the  brackets represents
the correction due  to the presence of the failed coherent backscattering. Importantly,  Eq.~(\ref{screen-scM}) includes both, the   limit  of infinite capacitance $C$  and the transparent case.   In the charge neutrality  limit  ($\tau =0$) the presence of the tunnel barriers does not drastically alter the conclusion drawn for the transparent case.  Indeed,  for the weak-localization correction,  in addition to the expected substitution  $N_{\al}, N$ by $\Gamma_{\al}^{(1)}, {\cal N}$ ,  we observe only a renormalisation by a factor $( { \Gamma_{\al}^{(2)} / \Gamma_{\al}^{(1)}  }   +  {  \Gamma_{\bet}^{(2)} /  \Gamma_{\bet}^{(1)}  } - \tilde{\cal N} / {\cal N})$. Thus Eq.~(\ref{screen-sc-CN})  becomes 
   \bea
&& 
\hspace{-0.5cm}
g^{{\rm wl}, \tau= 0}_{\al\bet}(\omega)
=
  \label{screen-sc-CN-tunel}
\\
&&
\le(
{ \Gamma_{\al}^{(2)}  \over \Gamma_{\al}^{(1)}  }   +
  {  \Gamma_{\bet}^{(2)} \over  \Gamma_{\bet}^{(1)}  } - {\tilde{\cal N} \over {\cal N} } \rg) 
\frac{
\Gamma_{\al}^{(1)}\Gamma_{\bet}^{(1)} 
}{
{\mathcal N} ^2
}
\frac{
e^{-\Theta_{\tE}} e^{  {\it i} \omega (\tEc+\tEo) }
  }{
 (1 - {\it i} \omega\tDI)  
 }.\nonumber
\eea

More importantly, one  of the  main effects of the   tunnel barrier in the dc-case  was the suppression of the
weak-localization correction~\cite{lid90,Whi07} for opaque barriers. This suppression results from the
competition between two purely quantum effects, interference and tunneling.  The corresponding semiclassical
treatment~\cite{Whi07} shows  that the cancellation  is  due to  an  exact compensation between the
weak-localization correction and  the failed coherent backscattering.    
It is interesting that  this conclusion cannot be  generalized to ac-transport.  
Since the frequency dependence of  the weak-localization correction differs from the one of  
the failed coherent backscattering the  compensation cannot occur.  Dynamical weak localization is thus more robust against the presence of tunnel barriers. We note, however, that for $\tau=0$ we recover the cancellation of the weak-localization correction with tunnel probabilities, see Eq.~(\ref{screen-sc-CN-tunel}).

\section{Charge relaxation resistance  of a mesoscopic chaotic capacitor}\label{mesocapa}

To illustrate and apply the general results derived above, we consider here the mesoscopic equivalent of a
classical RC circuit~\cite{But93a}.  A quantum  coherent capacitor  has been recently investigated experimentally  
by Gabelli et al.~\cite{Gab06a} using a two-dimensional electron gas.   The quantum capacitor is  composed of 
a macroscopic metallic electrode on top of a lateral quantum dot defining the second electrode. 
The role of the resistance is played by a quantum point contact that connects the quantum dot to a reservoir.  
The experiment  was  performed in the coherent regime at high magnetic field in the one edge state limit.   
Measuring the real and imaginary part  of the admittance of such a circuit, Ref.~[\onlinecite{Gab06a}]  confirmed the predicted~\cite{But93a} universal  value of the quantized charge relaxation resistance of a single channel cavity, which  is  equal to half a resistance quantum $h/2e^2$.

Based on this experimental realization we propose here to investigate the opposite regime  of large channel 
numbers  at zero magnetic field. This regime  is not characterized by the universal value of the preceding 
fully quantum  one,  however it should be  experimentally accessible.  If we assume that the  quantum dot is chaotic we  can map this system  to the one-terminal geometry of the more general set-up considered in the previous section.  The transparency of the quantum point contact is replaced by the transmission probability of the  tunnel barrier ${\bf \Gamma}_1$.  To simplify the result we assume in the following that the $N$ channels of the capacitor have the same tunnel rate, i.e.\  $\Gamma_{1,i} = \Gamma$  ($\forall i$), the dwell time of the capacitor is thus $\tD = \tH / (N\Gamma)$.

In a quantum coherent capacitor, there is obviously no dc-current, but we can address ac-transport via the 
admittance $G(\om)$~\cite{But93c,Gop96}.  At low temperatures it is characterized by an electrochemical 
capacitance $C_{\mu}$ and a charge relaxation resistance $R_{q}$,
\bea\label{Gexpand}
G(\om) = -{\it i} \om C_{\mu} + \om^2 C_{\mu}^2R_{q} + {\cal O} (\om^3) \, .
\eea
In contrast to their classical counterparts, $C_{\mu}$ and $R_{q}$ strongly depend on the local density inside the sample~\cite{But99}. They are thus sensitive to the phase coherent dynamics  of the electrons inside the sample and thus subject to dephasing.

\begin{figure}[h]
\begin{center}
\rotatebox{0}{\resizebox{8.5cm}{!}{\includegraphics{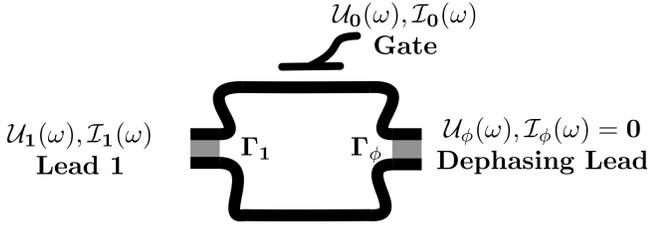}}}
\caption{Schematic picture of the mesoscopic capacitor with the dephasing lead (see text). The chaotic cavity has an extra lead (lead $\phi$), whose voltage is chosen to render the net current zero, which leads to dephasing without a loss of particles.  Since  each channel has the same tunnel rate $\Gamma_{1,i} = \Gamma$ and  $\Gamma_{\phi,i} = \Gamma_{\phi}$, $\forall i$, the dwell time of the capacitor is $\tD \propto (N\Gamma)^{-1} $ and the dephasing time reads $\tau_{\phi}\propto (N_{\phi}\Gamma_{\phi})^{-1} $.}
\label{fig4}
\end{center}
\end{figure}

To model  the loss of coherence of electrons inside the cavity we appeal to the so-called voltage/dephasing probe 
model~\cite{But86}, 
which consists of adding another lead $\phi$,  (see Fig.~\ref{fig4}) to our cavity and tuning the potential of this
probe in such a way that the net current is zero.  Consequently any electron that enters this lead is immediately
replaced by another one with an unrelated phase leading to incoherence without loss of particles.  While such an
approach has recently been used for the mesoscopic capacitor in the one channel limit~\cite{Nig07}, here we
investigate similar effects of the dephasing in the limit of large channel numbers~\cite{caveat}
where our semiclassical method is fully justified.

The  admittance can be  written as 
\begin{eqnarray} \label{gcapa}
G (\om ) = { -{\it i}\omega C \chi(\omega) \over -{\it i}\omega C + \chi(\omega) },
\end{eqnarray}
where
\begin{eqnarray}\label{chi}
 \chi(\omega)  =  G_{0}
\le(
g^{u}_{11}(\omega) - {  g^{u}_{1\phi}(\omega)  g^{u}_{\phi 1}(\omega) \over   g^{u}_{\phi\phi}(\omega)} 
\rg).
\end{eqnarray}

The unscreened admittance elements are  given in Eq.~(\ref{unscreen-scM}).  The
 survival times $\tDI$ and   $\tDII$ of this two-lead geometry are related to the real dwell time $\tD$ of our
 capacitor and to the dephasing time $\tau_{\phi} = \tH/(N_{\phi}\Gamma_{\phi})$, where $N_{\phi}$ and
 $\Gamma_{\phi,i} = \Gamma_{\phi}$ ($\forall i$) are, respectively, the number of channels and the tunneling rates of the dephasing lead~\cite{Pet07c}:  
\begin{subequations}
\bea\label{dephtime}
\tDI &=&  \tD\le[1+ {\tD\over \tau_{\phi}}\rg]^{-1}, \\
 \tDII&=& \tD
 \le[  (2 -\Gamma) +  {\tD\over \tau_{\phi} } (2 - \Gamma_{\phi} )\rg]^{-1}.
\eea
\end{subequations}
Inserting expression (\ref{unscreen-scM}) for the unscreened admittance elements  into Eqs.~(\ref{gcapa}, \ref{chi}) and performing an expansion in $\omega$ we get
\begin{subequations}
\begin{eqnarray}
C_{\mu} &=&  \frac{C e^2 \nu }{C +  e^2 \nu},\\
G_{0}R_{q} &=&  {1\over \Gamma  N }  + { D(\tEc,\tEo,\tau_{\phi}) \over (\Gamma  N)^2}  + {\cal O} (N^{-3}).
\end{eqnarray}
\end{subequations}
where we additionally used the relation between the mean  density of states, $\nu$, and the dwell time, 
$\tD= h\nu/ (d_sN\Gamma)$.  The dephasing function $D(\tEc,\tEo,\tau_{\phi}) $ reads
\be
D(\tEc,\tEo,\tau_{\phi}) =
 \Gamma        e^{ - { \tEo  \over \tD    } (1 -\Gamma)  - { \tEc \over \tD}   } \,\,
 \frac{
    e^{ - { \tEo  \over \tau_{\phi}    }  (1 -\Gamma_{\phi} ) - { \tEc \over \tau_{\phi} }} 
 }{
 \le(1 +  { \tD \over \tau_{\rm \phi}    }\rg)
 }.
 \ee

We finally consider the effect of a magnetic flux on the charge  relaxation resistance.
Substituting Eq.~(\ref{gwlmagneto}) (the dwell time being replaced by the survival time $\tDI$) into Eq.~(\ref{gcapa}) leaves the electrochemical  capacitance $C_{\mu}$ unchanged; only the dephasing function $D(\tEc,\tEo,\tau_{\phi}) $ is affected and replaced by
\be
D(\tEc,\tEo,\tau_{\phi},{ \Phi}) =
 \Gamma        e^{ - { \tEo (1 -\Gamma)   \over \tD    }  - { \tEc \over \tD}   } 
 \frac{
    e^{ - { \tEo (1 -\Gamma_{\phi} )  \over \tau_{\phi}    } - { \tEc \over \tau_{\phi} }} 
 }{
 \le(1 +  {\cal A}^2 { \Phi}^2 { \tD \over \tau_{\rm f}    } + { \tD \over \tau_{\phi}    }\rg)
 }.
 \ee

From this semiclassical investigation of the charge relaxation resistance, we can see that the fully coherent limit ($\tau_{\phi} =  \infty $,  $\tEo=\tEc =  0 $, ${\Phi} = 0$) delivers,  
\begin{eqnarray}\label{qrelax}
R_{q} =  \frac{1}{G_0 }  \frac{1}{\Gamma N  }
\le( 
1 +   \frac{1}{N  } 
\rg)
+ {\cal O} (N^{-3}).
\end{eqnarray}

Eq.~(\ref{qrelax}) is the first derivation of  the charge relaxation resitance in the large $N$ limit in presence of tunnel barriers. While  the leading order was guessed~\cite{Blan00},  the  weak-localization correction to $R_{q}$ has never been calculated before. Surprisingly,  it is
   linear in the inverse tunnel rate $\Gamma^{-1}$,  indicating  that the calculation of the  sub-leading  order correction  cannot be simply obtained  by an effective renormalisation of the channel number $N_{\rm eff} = \Gamma N $.  

For the incoherent limit,   obtained either by $\tE  \to  \infty $,  ${ \Phi}  \to  \infty$  or $\tau_{\phi} = 0$,
we get a suppression of the weak-localization correction  and thus  $R_{q} $ reduces to  
\begin{eqnarray}
R_{q} =  \frac{1}{G_0 }   
\frac{1}{\Gamma N}.
\end{eqnarray}

This  value corresponds to the fully incoherent limit that corresponds to the two-terminal resistance, and has been obtained under the simple application of our dephasing process.   Interestingly, this limit was not trivial to obtain in the edge state calculation~\cite{Nig07} ($N =1$), where  perfect inter-channel relaxation inside the voltage probe was assumed. This seems not to be required  in the fully chaotic case in the limit  $N \gg1$.

\section{Conclusions}

In this work we focused on the topic of ac-transport through chaotic ballistic conductors, addressing
in particular weak localization corrections to the admittance from a semiclassical perspective. 
Employing trajectory-based methods we confirmed RMT results for the bare and screened admittance and,
going beyond RMT, derived the Ehrenfest time dependence. The Ehrenfest timescale enters twice into 
the expressions for dynamical weak localization: first, as an exponential suppression with an exponent
given by the ratio of the Ehrenfest and dwell time, $\tE/\tD$; second the dynamical weak 
localization acquires an oscillatory frequency-dependent behavior of period  $2\tE$, which may be
amenable to measurements based on variations of the ac-frequency.
We emphasize that our results are valid for any finite capacitance $C$ and hence not limited to the  
electroneutrality assumption  of Ref.~[\onlinecite{Ale96}]. This extends the class of experimental settings
for which the Ehrenfest time correction can be investigated. More generally, the results presented
 underline, firstly, the power of semiclassical techniques to provide a clear and quantitative
picture of ac-driven quantum transport in the various regimes and, secondly, they give a
justification of the "stub model"\cite{Pol03} in the low-frequency regime.
 
Moreover we took into account tunnel barriers in the semiclassical approach to the ac-admittance,
extending the work of Whitney\cite{Whi07} on dc-transport. This led us to a general formulation of 
ac-transport. One main conclusion is that weak-localization is more robust against effects of 
tunnel barriers in the dynamical than in the dc-regime.   
The extension of our semiclassical treatment to tunnel barriers also enables us to access 
the experimentally relevant case of a quantum coherent capacitor, for which we provide the first 
derivation of the weak-localization correction to the charge relaxation resistance in presence of 
tunnel barriers. 
 
We add that, from a methodological point of view, the semiclassical approach presented might be
helpful to achieve a better understanding of the proximity effect on the density of states of 
chaotic Andreev billiards. Finally, the ac-conductance discussed here is closely related to 
problems of computing (photo-)absorption and, more generally, linear-response based dynamical 
susceptibilities for mesoscopic quantum systems. It appears promising to apply the
semiclassical techniques, developed here for (ac-)quantum transport, to refine earlier 
semiclassical approaches\cite{Meh98} to  (photo-)absorption in closed ballistic cavities
or metal clusters, which additionally poses the challenge to semiclassically cope with screening 
effects and plasmon excitations.

\section*{ACKNOWLEDGMENTS}

The authors thank P.W.~Brouwer, M.~B\"uttiker, M.~Guti\'errez, S.~Nigg,  M.~Polianski and R.S.~Whitney
for valuable and stimulating discussions.  We acknowledge funding from the DFG under GRK 638 
and from the Alexander von Humboldt foundation (C.~P. and J.~K.).


\begin{thebibliography}{76}
\expandafter\ifx\csname natexlab\endcsname\relax\def\natexlab#1{#1}\fi
\expandafter\ifx\csname bibnamefont\endcsname\relax
  \def\bibnamefont#1{#1}\fi
\expandafter\ifx\csname bibfnamefont\endcsname\relax
  \def\bibfnamefont#1{#1}\fi
\expandafter\ifx\csname citenamefont\endcsname\relax
  \def\citenamefont#1{#1}\fi
\expandafter\ifx\csname url\endcsname\relax
  \def\url#1{\texttt{#1}}\fi
\expandafter\ifx\csname urlprefix\endcsname\relax\def\urlprefix{URL }\fi
\providecommand{\bibinfo}[2]{#2}
\providecommand{\eprint}[2][]{\url{#2}}

\bibitem[{\citenamefont{Pieper and Price}(1994)}]{Pie94a}
\bibinfo{author}{\bibfnamefont{J.~B.} \bibnamefont{Pieper}} \bibnamefont{and}
  \bibinfo{author}{\bibfnamefont{J.~C.} \bibnamefont{Price}},
  \bibinfo{journal}{Phys. Rev. Lett.} \textbf{\bibinfo{volume}{72}},
  \bibinfo{pages}{3586} (\bibinfo{year}{1994}).

\bibitem[{\citenamefont{Chen et~al.}(1994)\citenamefont{Chen, Smith,
  B{\"u}ttiker, and Shayegan}}]{Che94}
\bibinfo{author}{\bibfnamefont{W.}~\bibnamefont{Chen}},
  \bibinfo{author}{\bibfnamefont{T.~P.} \bibnamefont{Smith}},
  \bibinfo{author}{\bibfnamefont{M.}~\bibnamefont{B{\"u}ttiker}},
  \bibnamefont{and} \bibinfo{author}{\bibfnamefont{M.}~\bibnamefont{Shayegan}},
  \bibinfo{journal}{Phys. Rev. Lett.} \textbf{\bibinfo{volume}{73}},
  \bibinfo{pages}{146} (\bibinfo{year}{1994}).

\bibitem[{\citenamefont{Kouwenhoven et~al.}(1994)\citenamefont{Kouwenhoven,
  Jauhar, Orenstein, McEuen, Nagamune, Motohisa, and Sakaki}}]{Kou94}
\bibinfo{author}{\bibfnamefont{L.~P.} \bibnamefont{Kouwenhoven}},
  \bibinfo{author}{\bibfnamefont{S.}~\bibnamefont{Jauhar}},
  \bibinfo{author}{\bibfnamefont{J.}~\bibnamefont{Orenstein}},
  \bibinfo{author}{\bibfnamefont{P.~L.} \bibnamefont{McEuen}},
  \bibinfo{author}{\bibfnamefont{Y.}~\bibnamefont{Nagamune}},
  \bibinfo{author}{\bibfnamefont{J.}~\bibnamefont{Motohisa}}, \bibnamefont{and}
  \bibinfo{author}{\bibfnamefont{H.}~\bibnamefont{Sakaki}},
  \bibinfo{journal}{Phys. Rev. Lett.} \textbf{\bibinfo{volume}{73}},
  \bibinfo{pages}{3443} (\bibinfo{year}{1994}).

\bibitem[{\citenamefont{Reznikov et~al.}(1995)\citenamefont{Reznikov, Heiblum,
  Shtrikman, and Mahalu}}]{Rez95}
\bibinfo{author}{\bibfnamefont{M.}~\bibnamefont{Reznikov}},
  \bibinfo{author}{\bibfnamefont{M.}~\bibnamefont{Heiblum}},
  \bibinfo{author}{\bibfnamefont{H.}~\bibnamefont{Shtrikman}},
  \bibnamefont{and} \bibinfo{author}{\bibfnamefont{D.}~\bibnamefont{Mahalu}},
  \bibinfo{journal}{Phys. Rev. Lett.} \textbf{\bibinfo{volume}{75}},
  \bibinfo{pages}{3340} (\bibinfo{year}{1995}).

\bibitem[{\citenamefont{Verghese et~al.}(1995)\citenamefont{Verghese, Wyss,
  F\"orster, Rooks, and Hu}}]{Ver95}
\bibinfo{author}{\bibfnamefont{S.}~\bibnamefont{Verghese}},
  \bibinfo{author}{\bibfnamefont{R.~A.} \bibnamefont{Wyss}},
  \bibinfo{author}{\bibfnamefont{A.}~\bibnamefont{F\"orster}},
  \bibinfo{author}{\bibfnamefont{M.~J.} \bibnamefont{Rooks}}, \bibnamefont{and}
  \bibinfo{author}{\bibfnamefont{Q.}~\bibnamefont{Hu}}, \bibinfo{journal}{Phys.
  Rev. B} \textbf{\bibinfo{volume}{52}}, \bibinfo{pages}{14834}
  (\bibinfo{year}{1995}).

\bibitem[{\citenamefont{Schoelkopf et~al.}(1997)\citenamefont{Schoelkopf,
  Burke, Kozhevnikov, Prober, and Rooks}}]{Scho97}
\bibinfo{author}{\bibfnamefont{R.}~\bibnamefont{Schoelkopf}},
  \bibinfo{author}{\bibfnamefont{P.}~\bibnamefont{Burke}},
  \bibinfo{author}{\bibfnamefont{A.}~\bibnamefont{Kozhevnikov}},
  \bibinfo{author}{\bibfnamefont{D.}~\bibnamefont{Prober}}, \bibnamefont{and}
  \bibinfo{author}{\bibfnamefont{M.}~\bibnamefont{Rooks}},
  \bibinfo{journal}{Phys. Rev. Lett.} \textbf{\bibinfo{volume}{78}},
  \bibinfo{pages}{3370} (\bibinfo{year}{1997}).

\bibitem[{\citenamefont{Reydellet et~al.}(2003)\citenamefont{Reydellet, Roche,
  Glattli, Etienne, and Jin}}]{Rey03}
\bibinfo{author}{\bibfnamefont{L.}~\bibnamefont{Reydellet}},
  \bibinfo{author}{\bibfnamefont{P.}~\bibnamefont{Roche}},
  \bibinfo{author}{\bibfnamefont{D.}~\bibnamefont{Glattli}},
  \bibinfo{author}{\bibfnamefont{B.}~\bibnamefont{Etienne}}, \bibnamefont{and}
  \bibinfo{author}{\bibfnamefont{Y.}~\bibnamefont{Jin}},
  \bibinfo{journal}{Phys. Rev. Lett.} \textbf{\bibinfo{volume}{90}},
  \bibinfo{pages}{176803} (\bibinfo{year}{2003}).

\bibitem[{\citenamefont{Gabelli et~al.}(2006)\citenamefont{Gabelli, F\`eve,
  Berroir, and Placais}}]{Gab06a}
\bibinfo{author}{\bibfnamefont{J.}~\bibnamefont{Gabelli}},
  \bibinfo{author}{\bibfnamefont{G.}~\bibnamefont{F\`eve}},
  \bibinfo{author}{\bibfnamefont{J.}~\bibnamefont{Berroir}}, \bibnamefont{and}
  \bibinfo{author}{\bibfnamefont{B.}~\bibnamefont{Placais}},
  \bibinfo{journal}{Science} \textbf{\bibinfo{volume}{313}},
  \bibinfo{pages}{499} (\bibinfo{year}{2006}).

\bibitem[{\citenamefont{F\`eve et~al.}(2007)\citenamefont{F\`eve, Mah\'e,
  Berroir, Kontos, Placais, Glattli, Cavanna, Etienne, and Jin}}]{Fev07}
\bibinfo{author}{\bibfnamefont{G.}~\bibnamefont{F\`eve}},
  \bibinfo{author}{\bibfnamefont{A.}~\bibnamefont{Mah\'e}},
  \bibinfo{author}{\bibfnamefont{J.}~\bibnamefont{Berroir}},
  \bibinfo{author}{\bibfnamefont{T.}~\bibnamefont{Kontos}},
  \bibinfo{author}{\bibfnamefont{B.}~\bibnamefont{Placais}},
  \bibinfo{author}{\bibfnamefont{D.}~\bibnamefont{Glattli}},
  \bibinfo{author}{\bibfnamefont{A.}~\bibnamefont{Cavanna}},
  \bibinfo{author}{\bibfnamefont{B.}~\bibnamefont{Etienne}}, \bibnamefont{and}
  \bibinfo{author}{\bibfnamefont{Y.}~\bibnamefont{Jin}},
  \bibinfo{journal}{Science} \textbf{\bibinfo{volume}{316}},
  \bibinfo{pages}{1169} (\bibinfo{year}{2007}).

\bibitem[{\citenamefont{Nagaev et~al.}(2004)\citenamefont{Nagaev, Pilgram, and
  B\"uttiker}}]{Nag04}
\bibinfo{author}{\bibfnamefont{K.}~\bibnamefont{Nagaev}},
  \bibinfo{author}{\bibfnamefont{S.}~\bibnamefont{Pilgram}}, \bibnamefont{and}
  \bibinfo{author}{\bibfnamefont{M.}~\bibnamefont{B\"uttiker}},
  \bibinfo{journal}{Phys. Rev. Lett.} \textbf{\bibinfo{volume}{92}},
  \bibinfo{pages}{176804} (\bibinfo{year}{2004}).

\bibitem[{\citenamefont{Hekking and Pekola}(2006)}]{Hek06}
\bibinfo{author}{\bibfnamefont{F.~W.~J.} \bibnamefont{Hekking}}
  \bibnamefont{and} \bibinfo{author}{\bibfnamefont{J.~P.}
  \bibnamefont{Pekola}}, \bibinfo{journal}{Phys. Rev. Lett.}
  \textbf{\bibinfo{volume}{96}}, \bibinfo{pages}{056603}
  (\bibinfo{year}{2006}).

\bibitem[{\citenamefont{Salo and Pekola}(2006)}]{Sal06}
\bibinfo{author}{\bibfnamefont{J.}~\bibnamefont{Salo}} \bibnamefont{and}
  \bibinfo{author}{\bibfnamefont{J.~P.} \bibnamefont{Pekola}},
  \bibinfo{journal}{Phys. Rev. B} \textbf{\bibinfo{volume}{74}},
  \bibinfo{pages}{125427} (\bibinfo{year}{2006}).

\bibitem[{\citenamefont{Bagrets and Pistolesi}(2007)}]{Bag07}
\bibinfo{author}{\bibfnamefont{D.}~\bibnamefont{Bagrets}} \bibnamefont{and}
  \bibinfo{author}{\bibfnamefont{F.}~\bibnamefont{Pistolesi}},
  \bibinfo{journal}{Phys. Rev. B} \textbf{\bibinfo{volume}{75}},
  \bibinfo{pages}{165315} (\bibinfo{year}{2007}).

\bibitem[{\citenamefont{Nigg and B\"uttiker}(2008)}]{Nig07}
\bibinfo{author}{\bibfnamefont{S.}~\bibnamefont{Nigg}} \bibnamefont{and}
  \bibinfo{author}{\bibfnamefont{M.}~\bibnamefont{B\"uttiker}},
  \bibinfo{journal}{Phys. Rev. B} \textbf{\bibinfo{volume}{77}},
  \bibinfo{pages}{085312} (\bibinfo{year}{2008}).

\bibitem[{\citenamefont{Moskalets et~al.}(2008)\citenamefont{Moskalets,
  Samuelsson, and B\"uttiker}}]{Mos08}
\bibinfo{author}{\bibfnamefont{M.}~\bibnamefont{Moskalets}},
  \bibinfo{author}{\bibfnamefont{P.}~\bibnamefont{Samuelsson}},
  \bibnamefont{and}
  \bibinfo{author}{\bibfnamefont{M.}~\bibnamefont{B\"uttiker}},
  \bibinfo{journal}{Phys. Rev. Lett.} \textbf{\bibinfo{volume}{100}},
  \bibinfo{pages}{086601} (\bibinfo{year}{2008}).

\bibitem[{\citenamefont{Safi et~al.}(2008)\citenamefont{Safi, Bena, and
  Cr{\'e}pieux}}]{Saf08}
\bibinfo{author}{\bibfnamefont{I.}~\bibnamefont{Safi}},
  \bibinfo{author}{\bibfnamefont{C.}~\bibnamefont{Bena}}, \bibnamefont{and}
  \bibinfo{author}{\bibfnamefont{A.}~\bibnamefont{Cr{\'e}pieux}},
  \bibinfo{journal}{Phys. Rev. B} \textbf{\bibinfo{volume}{78}},
  \bibinfo{pages}{205422} (\bibinfo{year}{2008}).

\bibitem[{\citenamefont{Park and Ahn}(2008)}]{Par08}
\bibinfo{author}{\bibfnamefont{H.~C.} \bibnamefont{Park}} \bibnamefont{and}
  \bibinfo{author}{\bibfnamefont{K.-H.} \bibnamefont{Ahn}},
  \bibinfo{journal}{Phys. Rev. Lett.} \textbf{\bibinfo{volume}{101}},
  \bibinfo{pages}{116804} (\bibinfo{year}{2008}).

\bibitem[{\citenamefont{Fisher and Lee}(1981)}]{Fis81}
\bibinfo{author}{\bibfnamefont{D.}~\bibnamefont{Fisher}} \bibnamefont{and}
  \bibinfo{author}{\bibfnamefont{P.}~\bibnamefont{Lee}},
  \bibinfo{journal}{Phys. Rev. B} \textbf{\bibinfo{volume}{23}},
  \bibinfo{pages}{6851} (\bibinfo{year}{1981}).

\bibitem[{\citenamefont{Baranger and Stone}(1989)}]{Bar89}
\bibinfo{author}{\bibfnamefont{H.~U.} \bibnamefont{Baranger}} \bibnamefont{and}
  \bibinfo{author}{\bibfnamefont{A.~D.} \bibnamefont{Stone}},
  \bibinfo{journal}{Phys. Rev. B} \textbf{\bibinfo{volume}{40}},
  \bibinfo{pages}{8169} (\bibinfo{year}{1989}).

\bibitem[{\citenamefont{Shepard}(1991)}]{She91}
\bibinfo{author}{\bibfnamefont{K.}~\bibnamefont{Shepard}},
  \bibinfo{journal}{Phys. Rev. B} \textbf{\bibinfo{volume}{43}},
  \bibinfo{pages}{11623} (\bibinfo{year}{1991}).

\bibitem[{\citenamefont{Pastawski}(1991)}]{Pas91}
\bibinfo{author}{\bibfnamefont{H.~M.} \bibnamefont{Pastawski}},
  \bibinfo{journal}{Phys. Rev. B} \textbf{\bibinfo{volume}{44}},
  \bibinfo{pages}{6329} (\bibinfo{year}{1991}).

\bibitem[{\citenamefont{B\"uttiker et~al.}(1993)\citenamefont{B\"uttiker,
  Pr\^etre, and Thomas}}]{But93a}
\bibinfo{author}{\bibfnamefont{M.}~\bibnamefont{B\"uttiker}},
  \bibinfo{author}{\bibfnamefont{A.}~\bibnamefont{Pr\^etre}}, \bibnamefont{and}
  \bibinfo{author}{\bibfnamefont{H.}~\bibnamefont{Thomas}},
  \bibinfo{journal}{Phys. Rev. Lett.} \textbf{\bibinfo{volume}{70}},
  \bibinfo{pages}{4114} (\bibinfo{year}{1993}).

\bibitem[{\citenamefont{B\"uttiker}(1993)}]{But93c}
\bibinfo{author}{\bibfnamefont{M.}~\bibnamefont{B\"uttiker}},
  \bibinfo{journal}{J. Phys. C} \textbf{\bibinfo{volume}{5}},
  \bibinfo{pages}{9361} (\bibinfo{year}{1993}).

\bibitem[{\citenamefont{Vavilov}(2005)}]{Vav05}
\bibinfo{author}{\bibfnamefont{M.~G.} \bibnamefont{Vavilov}},
  \bibinfo{journal}{J. Phys. A} \textbf{\bibinfo{volume}{38}},
  \bibinfo{pages}{10587} (\bibinfo{year}{2005}).

\bibitem[{\citenamefont{Polianski and Brouwer}(2003)}]{Pol03}
\bibinfo{author}{\bibfnamefont{M.}~\bibnamefont{Polianski}} \bibnamefont{and}
  \bibinfo{author}{\bibfnamefont{P.~W.} \bibnamefont{Brouwer}},
  \bibinfo{journal}{J. Phys. A} \textbf{\bibinfo{volume}{36}},
  \bibinfo{pages}{3215} (\bibinfo{year}{2003}).

\bibitem[{\citenamefont{B\"uttiker and Christen}(1997)}]{But97}
\bibinfo{author}{\bibfnamefont{M.}~\bibnamefont{B\"uttiker}} \bibnamefont{and}
  \bibinfo{author}{\bibfnamefont{T.}~\bibnamefont{Christen}}, in
  \emph{\bibinfo{booktitle}{Mesoscopic Electron Transport}}, edited by
  \bibinfo{editor}{\bibfnamefont{L.}~\bibnamefont{Kowenhoven}},
  \bibinfo{editor}{\bibfnamefont{G.}~\bibnamefont{Sch\"on}}, \bibnamefont{and}
  \bibinfo{editor}{\bibfnamefont{L.}~\bibnamefont{Sohn}}
  (\bibinfo{publisher}{NATO ASI Series E, Kluwer Ac. Publ., Dordrecht},
  \bibinfo{year}{1997}), p. \bibinfo{pages}{259}.

\bibitem[{not()}]{note1}
\bibinfo{note}{The regime of larger ac-driving frequency requires other
  techniques such as Floquet scattering theory which has been applied in the
  context of ac-driven chaotic scattering, e.g. in Ref.~[\onlinecite{Hen01}].}

\bibitem[{\citenamefont{Bl\"umel and Smilansky}(1988)}]{Blu88}
\bibinfo{author}{\bibfnamefont{R.}~\bibnamefont{Bl\"umel}} \bibnamefont{and}
  \bibinfo{author}{\bibfnamefont{U.}~\bibnamefont{Smilansky}},
  \bibinfo{journal}{Phys. Rev. Lett.} \textbf{\bibinfo{volume}{60}},
  \bibinfo{pages}{477} (\bibinfo{year}{1988}).

\bibitem[{\citenamefont{Baranger et~al.}(1993)\citenamefont{Baranger, Jalabert,
  and Stone}}]{Bar93a}
\bibinfo{author}{\bibfnamefont{H.~U.} \bibnamefont{Baranger}},
  \bibinfo{author}{\bibfnamefont{R.~A.} \bibnamefont{Jalabert}},
  \bibnamefont{and} \bibinfo{author}{\bibfnamefont{A.~D.} \bibnamefont{Stone}},
  \bibinfo{journal}{Phys. Rev. Lett.} \textbf{\bibinfo{volume}{70}},
  \bibinfo{pages}{3876} (\bibinfo{year}{1993}).

\bibitem[{\citenamefont{Verbaarschot et~al.}(1985)\citenamefont{Verbaarschot,
  Weidenm\"uller, and Zirnbauer}}]{Ver85}
\bibinfo{author}{\bibfnamefont{J.}~\bibnamefont{Verbaarschot}},
  \bibinfo{author}{\bibfnamefont{H.}~\bibnamefont{Weidenm\"uller}},
  \bibnamefont{and}
  \bibinfo{author}{\bibfnamefont{M.}~\bibnamefont{Zirnbauer}},
  \bibinfo{journal}{Phys. Rep.} \textbf{\bibinfo{volume}{129}},
  \bibinfo{pages}{367} (\bibinfo{year}{1985}).

\bibitem[{\citenamefont{Frahm}(1995)}]{Fra95}
\bibinfo{author}{\bibfnamefont{K.}~\bibnamefont{Frahm}}, \bibinfo{journal}{EPL}
  \textbf{\bibinfo{volume}{30}}, \bibinfo{pages}{457} (\bibinfo{year}{1995}).

\bibitem[{\citenamefont{Brouwer and B\"uttiker}(1997)}]{Brou97a}
\bibinfo{author}{\bibfnamefont{P.~W.} \bibnamefont{Brouwer}} \bibnamefont{and}
  \bibinfo{author}{\bibfnamefont{M.}~\bibnamefont{B\"uttiker}},
  \bibinfo{journal}{EPL} \textbf{\bibinfo{volume}{37}}, \bibinfo{pages}{441}
  (\bibinfo{year}{1997}).

\bibitem[{\citenamefont{Aleiner and Larkin}(1996)}]{Ale96}
\bibinfo{author}{\bibfnamefont{I.}~\bibnamefont{Aleiner}} \bibnamefont{and}
  \bibinfo{author}{\bibfnamefont{A.}~\bibnamefont{Larkin}},
  \bibinfo{journal}{Phys. Rev. B} \textbf{\bibinfo{volume}{54}},
  \bibinfo{pages}{14423} (\bibinfo{year}{1996}).

\bibitem[{\citenamefont{Larkin and Ovchinnikov}(1969)}]{Lar69}
\bibinfo{author}{\bibfnamefont{A.}~\bibnamefont{Larkin}} \bibnamefont{and}
  \bibinfo{author}{\bibfnamefont{Y.}~\bibnamefont{Ovchinnikov}},
  \bibinfo{journal}{Sov. Phys. JETP} \textbf{\bibinfo{volume}{28}},
  \bibinfo{pages}{1200} (\bibinfo{year}{1969}).

\bibitem[{\citenamefont{Berman and Zaslavsky}(1978)}]{Berm79}
\bibinfo{author}{\bibfnamefont{G.}~\bibnamefont{Berman}} \bibnamefont{and}
  \bibinfo{author}{\bibfnamefont{G.}~\bibnamefont{Zaslavsky}},
  \bibinfo{journal}{Physica A} \textbf{\bibinfo{volume}{91A}},
  \bibinfo{pages}{450} (\bibinfo{year}{1978}).

\bibitem[{\citenamefont{Vavilov and Larkin}(2003)}]{Vav03}
\bibinfo{author}{\bibfnamefont{M.~G.} \bibnamefont{Vavilov}} \bibnamefont{and}
  \bibinfo{author}{\bibfnamefont{A.}~\bibnamefont{Larkin}},
  \bibinfo{journal}{Phys. Rev. B} \textbf{\bibinfo{volume}{67}},
  \bibinfo{pages}{115335} (\bibinfo{year}{2003}).

\bibitem[{\citenamefont{Schomerus and Jacquod}(2005)}]{Sch05}
\bibinfo{author}{\bibfnamefont{H.}~\bibnamefont{Schomerus}} \bibnamefont{and}
  \bibinfo{author}{\bibfnamefont{P.}~\bibnamefont{Jacquod}},
  \bibinfo{journal}{J. Phys. A} \textbf{\bibinfo{volume}{38}},
  \bibinfo{pages}{10663} (\bibinfo{year}{2005}).

\bibitem[{Wal()}]{Walbook}
\bibinfo{note}{For an introduction to the semiclassical techniques, see, e.g.
  D. Waltner and K. Richter in {\it Handbook of Nonlinear Dynamics in
  Nanosystems}, G. Radons et al. (Eds.), Wiley VCH, in press (2009).}

\bibitem[{\citenamefont{Richter and Sieber}(2002)}]{Ric02}
\bibinfo{author}{\bibfnamefont{K.}~\bibnamefont{Richter}} \bibnamefont{and}
  \bibinfo{author}{\bibfnamefont{M.}~\bibnamefont{Sieber}},
  \bibinfo{journal}{Phys. Rev. Lett.} \textbf{\bibinfo{volume}{89}},
  \bibinfo{pages}{206801} (\bibinfo{year}{2002}).

\bibitem[{\citenamefont{Adagideli}(2003)}]{Ada03}
\bibinfo{author}{\bibfnamefont{I.}~\bibnamefont{Adagideli}},
  \bibinfo{journal}{Phys. Rev. B} \textbf{\bibinfo{volume}{68}},
  \bibinfo{pages}{23308} (\bibinfo{year}{2003}).

\bibitem[{\citenamefont{M\"uller et~al.}(2007)\citenamefont{M\"uller, Heusler,
  Braun, and Haake}}]{Mul07}
\bibinfo{author}{\bibfnamefont{S.}~\bibnamefont{M\"uller}},
  \bibinfo{author}{\bibfnamefont{S.}~\bibnamefont{Heusler}},
  \bibinfo{author}{\bibfnamefont{P.}~\bibnamefont{Braun}}, \bibnamefont{and}
  \bibinfo{author}{\bibfnamefont{F.}~\bibnamefont{Haake}},
  \bibinfo{journal}{New J. Phys} \textbf{\bibinfo{volume}{9}},
  \bibinfo{pages}{12} (\bibinfo{year}{2007}).

\bibitem[{\citenamefont{Jacquod and Whitney}(2006)}]{Jac06a}
\bibinfo{author}{\bibfnamefont{P.}~\bibnamefont{Jacquod}} \bibnamefont{and}
  \bibinfo{author}{\bibfnamefont{R.}~\bibnamefont{Whitney}},
  \bibinfo{journal}{Phys. Rev. B} \textbf{\bibinfo{volume}{73}},
  \bibinfo{pages}{195115} (\bibinfo{year}{2006}).

\bibitem[{\citenamefont{Brouwer and Rahav}(2006)}]{Brou06c}
\bibinfo{author}{\bibfnamefont{P.~W.} \bibnamefont{Brouwer}} \bibnamefont{and}
  \bibinfo{author}{\bibfnamefont{S.}~\bibnamefont{Rahav}},
  \bibinfo{journal}{Phys. Rev. B} \textbf{\bibinfo{volume}{74}},
  \bibinfo{pages}{075322} (\bibinfo{year}{2006}).

\bibitem[{\citenamefont{Brouwer}(2007)}]{Brou07c}
\bibinfo{author}{\bibfnamefont{P.~W.} \bibnamefont{Brouwer}},
  \bibinfo{journal}{Phys. Rev. B} \textbf{\bibinfo{volume}{76}},
  \bibinfo{pages}{165313} (\bibinfo{year}{2007}).

\bibitem[{\citenamefont{Whitney}(2007)}]{Whi07}
\bibinfo{author}{\bibfnamefont{R.}~\bibnamefont{Whitney}},
  \bibinfo{journal}{Phys. Rev. B} \textbf{\bibinfo{volume}{75}},
  \bibinfo{pages}{235404} (\bibinfo{year}{2007}).

\bibitem[{\citenamefont{Petitjean et~al.}(2007)\citenamefont{Petitjean,
  Jacquod, and Whitney}}]{Pet07c}
\bibinfo{author}{\bibfnamefont{C.}~\bibnamefont{Petitjean}},
  \bibinfo{author}{\bibfnamefont{P.}~\bibnamefont{Jacquod}}, \bibnamefont{and}
  \bibinfo{author}{\bibfnamefont{R.}~\bibnamefont{Whitney}},
  \bibinfo{journal}{JETP. Lett.} \textbf{\bibinfo{volume}{86}},
  \bibinfo{pages}{736} (\bibinfo{year}{2007}).

\bibitem[{\citenamefont{Whitney et~al.}(2008)\citenamefont{Whitney, Jacquod,
  and Petitjean}}]{Whi08}
\bibinfo{author}{\bibfnamefont{R.~S.} \bibnamefont{Whitney}},
  \bibinfo{author}{\bibfnamefont{P.}~\bibnamefont{Jacquod}}, \bibnamefont{and}
  \bibinfo{author}{\bibfnamefont{C.}~\bibnamefont{Petitjean}},
  \bibinfo{journal}{Phys. Rev. B} \textbf{\bibinfo{volume}{77}},
  \bibinfo{pages}{045315} (\bibinfo{year}{2008}).

\bibitem[{\citenamefont{Kuipers and Sieber}(2008)}]{Kui08}
\bibinfo{author}{\bibfnamefont{J.}~\bibnamefont{Kuipers}} \bibnamefont{and}
  \bibinfo{author}{\bibfnamefont{M.}~\bibnamefont{Sieber}},
  \bibinfo{journal}{Phys. Rev. E} \textbf{\bibinfo{volume}{77}},
  \bibinfo{pages}{046219} (\bibinfo{year}{2008}).

\bibitem[{\citenamefont{Brouwer and Altland}(2008)}]{Brou08}
\bibinfo{author}{\bibfnamefont{P.~W.} \bibnamefont{Brouwer}} \bibnamefont{and}
  \bibinfo{author}{\bibfnamefont{A.}~\bibnamefont{Altland}},
  \bibinfo{journal}{Phys. Rev. B} \textbf{\bibinfo{volume}{78}},
  \bibinfo{pages}{075304} (\bibinfo{year}{2008}).

\bibitem[{\citenamefont{Wilkinson}(1987)}]{Wil87}
\bibinfo{author}{\bibfnamefont{M.}~\bibnamefont{Wilkinson}},
  \bibinfo{journal}{J. Phys. A} \textbf{\bibinfo{volume}{20}},
  \bibinfo{pages}{2415} (\bibinfo{year}{1987}).

\bibitem[{\citenamefont{Mehlig and Richter}(1998)}]{Meh98}
\bibinfo{author}{\bibfnamefont{B.}~\bibnamefont{Mehlig}} \bibnamefont{and}
  \bibinfo{author}{\bibfnamefont{K.}~\bibnamefont{Richter}},
  \bibinfo{journal}{Phys. Rev. Lett.} \textbf{\bibinfo{volume}{80}},
  \bibinfo{pages}{1936} (\bibinfo{year}{1998}).

\bibitem[{\citenamefont{Richter}(2000)}]{Ric00}
\bibinfo{author}{\bibfnamefont{K.}~\bibnamefont{Richter}},
  \emph{\bibinfo{title}{Semiclassical Theory of Mesoscopic Quantum Systems,
  Springer Tracts in Modern Physics Vol.~161}} (\bibinfo{publisher}{Springer,
  Berlin}, \bibinfo{year}{2000}).

\bibitem[{\citenamefont{Gopar et~al.}(1996)\citenamefont{Gopar, Mello, and
  B\"uttiker}}]{Gop96}
\bibinfo{author}{\bibfnamefont{V.}~\bibnamefont{Gopar}},
  \bibinfo{author}{\bibfnamefont{P.}~\bibnamefont{Mello}}, \bibnamefont{and}
  \bibinfo{author}{\bibfnamefont{M.}~\bibnamefont{B\"uttiker}},
  \bibinfo{journal}{Phys. Rev. Lett.} \textbf{\bibinfo{volume}{77}},
  \bibinfo{pages}{3005} (\bibinfo{year}{1996}).

\bibitem[{\citenamefont{Blanter and B\"uttiker}(2000)}]{Blan00}
\bibinfo{author}{\bibfnamefont{Y.}~\bibnamefont{Blanter}} \bibnamefont{and}
  \bibinfo{author}{\bibfnamefont{M.}~\bibnamefont{B\"uttiker}},
  \bibinfo{journal}{Phys. Rep.} \textbf{\bibinfo{volume}{336}}
  (\bibinfo{year}{2000}).

\bibitem[{\citenamefont{Pr{\^e}tre et~al.}(1996)\citenamefont{Pr{\^e}tre,
  Thomas, and B{\"u}ttiker}}]{Pre96}
\bibinfo{author}{\bibfnamefont{A.}~\bibnamefont{Pr{\^e}tre}},
  \bibinfo{author}{\bibfnamefont{H.}~\bibnamefont{Thomas}}, \bibnamefont{and}
  \bibinfo{author}{\bibfnamefont{M.}~\bibnamefont{B{\"u}ttiker}},
  \bibinfo{journal}{Phys. Rev. B} \textbf{\bibinfo{volume}{54}},
  \bibinfo{pages}{8130} (\bibinfo{year}{1996}).

\bibitem[{\citenamefont{Wang et~al.}(1997)\citenamefont{Wang, Zheng, and
  Guo}}]{Wan97}
\bibinfo{author}{\bibfnamefont{J.}~\bibnamefont{Wang}},
  \bibinfo{author}{\bibfnamefont{Q.}~\bibnamefont{Zheng}}, \bibnamefont{and}
  \bibinfo{author}{\bibfnamefont{H.}~\bibnamefont{Guo}},
  \bibinfo{journal}{Phys. Rev. B} \textbf{\bibinfo{volume}{55}},
  \bibinfo{pages}{9770} (\bibinfo{year}{1997}).

\bibitem[{\citenamefont{B\"uttiker}(2000)}]{But00}
\bibinfo{author}{\bibfnamefont{M.}~\bibnamefont{B\"uttiker}},
  \bibinfo{journal}{J. Low Temp.} \textbf{\bibinfo{volume}{118}},
  \bibinfo{pages}{519} (\bibinfo{year}{2000}).

\bibitem[{\citenamefont{Aleiner et~al.}(2002)\citenamefont{Aleiner, Brouwer,
  and Glazman}}]{Ale02}
\bibinfo{author}{\bibfnamefont{I.}~\bibnamefont{Aleiner}},
  \bibinfo{author}{\bibfnamefont{P.}~\bibnamefont{Brouwer}}, \bibnamefont{and}
  \bibinfo{author}{\bibfnamefont{L.}~\bibnamefont{Glazman}},
  \bibinfo{journal}{Phys. Rep.} \textbf{\bibinfo{volume}{358}},
  \bibinfo{pages}{309} (\bibinfo{year}{2002}).

\bibitem[{\citenamefont{Brouwer et~al.}(2005)\citenamefont{Brouwer, Lamacraft,
  and Flensberg}}]{Brou05b}
\bibinfo{author}{\bibfnamefont{P.}~\bibnamefont{Brouwer}},
  \bibinfo{author}{\bibfnamefont{A.}~\bibnamefont{Lamacraft}},
  \bibnamefont{and}
  \bibinfo{author}{\bibfnamefont{K.}~\bibnamefont{Flensberg}},
  \bibinfo{journal}{Phys. Rev. B} \textbf{\bibinfo{volume}{72}},
  \bibinfo{pages}{075316} (\bibinfo{year}{2005}).

\bibitem[{\citenamefont{Wang et~al.}(1999)\citenamefont{Wang, Wang, and
  Guo}}]{Wan99}
\bibinfo{author}{\bibfnamefont{B.}~\bibnamefont{Wang}},
  \bibinfo{author}{\bibfnamefont{J.}~\bibnamefont{Wang}}, \bibnamefont{and}
  \bibinfo{author}{\bibfnamefont{H.}~\bibnamefont{Guo}},
  \bibinfo{journal}{Phys. Rev. Lett.} \textbf{\bibinfo{volume}{82}},
  \bibinfo{pages}{398} (\bibinfo{year}{1999}).

\bibitem[{\citenamefont{Miller}(1975)}]{Mil75}
\bibinfo{author}{\bibfnamefont{W.}~\bibnamefont{Miller}},
  \bibinfo{journal}{Advances in Chemical Physics}
  \textbf{\bibinfo{volume}{30}}, \bibinfo{pages}{77} (\bibinfo{year}{1975}).

\bibitem[{\citenamefont{Gutzwiller}(1990)}]{Gut90}
\bibinfo{author}{\bibfnamefont{M.~C.} \bibnamefont{Gutzwiller}},
  \emph{\bibinfo{title}{Chaos in Classical and Quantum Mechanics}}
  (\bibinfo{publisher}{Springer, New York}, \bibinfo{year}{1990}).

\bibitem[{\citenamefont{Sieber}(1999)}]{Sie99b}
\bibinfo{author}{\bibfnamefont{M.}~\bibnamefont{Sieber}}, \bibinfo{journal}{J.
  Phys. A} \textbf{\bibinfo{volume}{32}}, \bibinfo{pages}{7679}
  (\bibinfo{year}{1999}).

\bibitem[{cav({\natexlab{a}})}]{caveat2}
\bibinfo{note}{Part IV "Time dependence" of Ref.~[\onlinecite{Brou06c}]. In
  order to compare with our result, one should at first neglect the difference
  between $\tEc$ and $\tEo$, then you substitute $\tau_{\rm abs}$ by $-{\it i}
  \om$ in Eq.~(52) and Eq.~(53) of Ref.~[\onlinecite{Brou06c}].}

\bibitem[{\citenamefont{Waltner et~al.}(2008)\citenamefont{Waltner,
  Guti\'errez, Goussev, and Richter}}]{Wal08}
\bibinfo{author}{\bibfnamefont{D.}~\bibnamefont{Waltner}},
  \bibinfo{author}{\bibfnamefont{M.}~\bibnamefont{Guti\'errez}},
  \bibinfo{author}{\bibfnamefont{A.}~\bibnamefont{Goussev}}, \bibnamefont{and}
  \bibinfo{author}{\bibfnamefont{K.}~\bibnamefont{Richter}},
  \bibinfo{journal}{Phys. Rev. Lett.} \textbf{\bibinfo{volume}{101}},
  \bibinfo{pages}{174101} (\bibinfo{year}{2008}).

\bibitem[{\citenamefont{Guti\'errez et~al.}(2009)\citenamefont{Guti\'errez,
  Waltner, Kuipers, and Richter}}]{Gut08}
\bibinfo{author}{\bibfnamefont{M.}~\bibnamefont{Guti\'errez}},
  \bibinfo{author}{\bibfnamefont{D.}~\bibnamefont{Waltner}},
  \bibinfo{author}{\bibfnamefont{J.}~\bibnamefont{Kuipers}}, \bibnamefont{and}
  \bibinfo{author}{\bibfnamefont{K.}~\bibnamefont{Richter}},
  \bibinfo{journal}{Phys. Rev. E} \textbf{\bibinfo{volume}{77}},
  \bibinfo{pages}{046212} (\bibinfo{year}{2009}).

\bibitem[{cav({\natexlab{b}})}]{caveat3}
\bibinfo{note}{As previously fixed in the dc-case see
  Refs.~[\onlinecite{Brou06c,Jac06a}], Ref.~[\onlinecite{Ale96}] had
  incorrectly treated the classical correlation at the encounter and therefore
  presented a discrepancy of a factor $2$ concerning the ratio $\tEc/\tD$.}

\bibitem[{\citenamefont{Oberholzer et~al.}(2002)\citenamefont{Oberholzer,
  Sukhorukov, and Sch\"onenberger}}]{Obe02}
\bibinfo{author}{\bibfnamefont{S.}~\bibnamefont{Oberholzer}},
  \bibinfo{author}{\bibfnamefont{E.}~\bibnamefont{Sukhorukov}},
  \bibnamefont{and}
  \bibinfo{author}{\bibfnamefont{C.}~\bibnamefont{Sch\"onenberger}},
  \bibinfo{journal}{Nature} \textbf{\bibinfo{volume}{415}},
  \bibinfo{pages}{765} (\bibinfo{year}{2002}).

\bibitem[{\citenamefont{Yevtushenko et~al.}(2000)\citenamefont{Yevtushenko,
  L\"utjering, Weiss, and Richter}}]{Yev00}
\bibinfo{author}{\bibfnamefont{O.}~\bibnamefont{Yevtushenko}},
  \bibinfo{author}{\bibfnamefont{G.}~\bibnamefont{L\"utjering}},
  \bibinfo{author}{\bibfnamefont{D.}~\bibnamefont{Weiss}}, \bibnamefont{and}
  \bibinfo{author}{\bibfnamefont{K.}~\bibnamefont{Richter}},
  \bibinfo{journal}{Phys. Rev. Lett.} \textbf{\bibinfo{volume}{84}},
  \bibinfo{pages}{542} (\bibinfo{year}{2000}).

\bibitem[{\citenamefont{Casati et~al.}(1997)\citenamefont{Casati, Maspero, and
  Shepelyansky}}]{Cas97}
\bibinfo{author}{\bibfnamefont{G.}~\bibnamefont{Casati}},
  \bibinfo{author}{\bibfnamefont{G.}~\bibnamefont{Maspero}}, \bibnamefont{and}
  \bibinfo{author}{\bibfnamefont{D.~L.} \bibnamefont{Shepelyansky}},
  \bibinfo{journal}{Phys. Rev. E} \textbf{\bibinfo{volume}{56}},
  \bibinfo{pages}{R6233} (\bibinfo{year}{1997}).

\bibitem[{\citenamefont{Frahm}(1997)}]{Fra97}
\bibinfo{author}{\bibfnamefont{K.~M.} \bibnamefont{Frahm}},
  \bibinfo{journal}{Phys. Rev. E} \textbf{\bibinfo{volume}{56}},
  \bibinfo{pages}{R6237} (\bibinfo{year}{1997}).

\bibitem[{\citenamefont{Iida et~al.}(1990)\citenamefont{Iida, Weidenm\"uller,
  and Zuk}}]{lid90}
\bibinfo{author}{\bibfnamefont{S.}~\bibnamefont{Iida}},
  \bibinfo{author}{\bibfnamefont{H.}~\bibnamefont{Weidenm\"uller}},
  \bibnamefont{and} \bibinfo{author}{\bibfnamefont{J.}~\bibnamefont{Zuk}},
  \bibinfo{journal}{Ann. Phys.} \textbf{\bibinfo{volume}{200}},
  \bibinfo{pages}{219} (\bibinfo{year}{1990}).

\bibitem[{\citenamefont{B\"uttiker}(1999)}]{But99}
\bibinfo{author}{\bibfnamefont{M.}~\bibnamefont{B\"uttiker}},
  \bibinfo{journal}{J. Korean Phys. Soc.} \textbf{\bibinfo{volume}{34}},
  \bibinfo{pages}{121} (\bibinfo{year}{1999}).

\bibitem[{\citenamefont{B\"uttiker}(1986)}]{But86}
\bibinfo{author}{\bibfnamefont{M.}~\bibnamefont{B\"uttiker}},
  \bibinfo{journal}{Phys. Rev. B} \textbf{\bibinfo{volume}{33}},
  \bibinfo{pages}{3020} (\bibinfo{year}{1986}).

\bibitem[{cav({\natexlab{c}})}]{caveat}
\bibinfo{note}{Strictly speaking, dephasing and voltage probe models are fully
  equivalent in the one channel limit only. However, they also lead to the same
  results in the limit of low temperature and frequency. See H. F\"orster, P.
  Samuelsson, S. Pilgram, and M. B\"uttiker, Phys. Rev. B {\bf 75}, 035340
  (2007)}.

\bibitem[{\citenamefont{Henseler et~al.}(2001)\citenamefont{Henseler, Dittrich,
  and Richter}}]{Hen01}
\bibinfo{author}{\bibfnamefont{M.}~\bibnamefont{Henseler}},
  \bibinfo{author}{\bibfnamefont{T.}~\bibnamefont{Dittrich}}, \bibnamefont{and}
  \bibinfo{author}{\bibfnamefont{K.}~\bibnamefont{Richter}},
  \bibinfo{journal}{Phys. Rev. E} \textbf{\bibinfo{volume}{64}},
  \bibinfo{pages}{046218} (\bibinfo{year}{2001}).

\end{thebibliography}
\end{document}